SEPTEMBER 2025

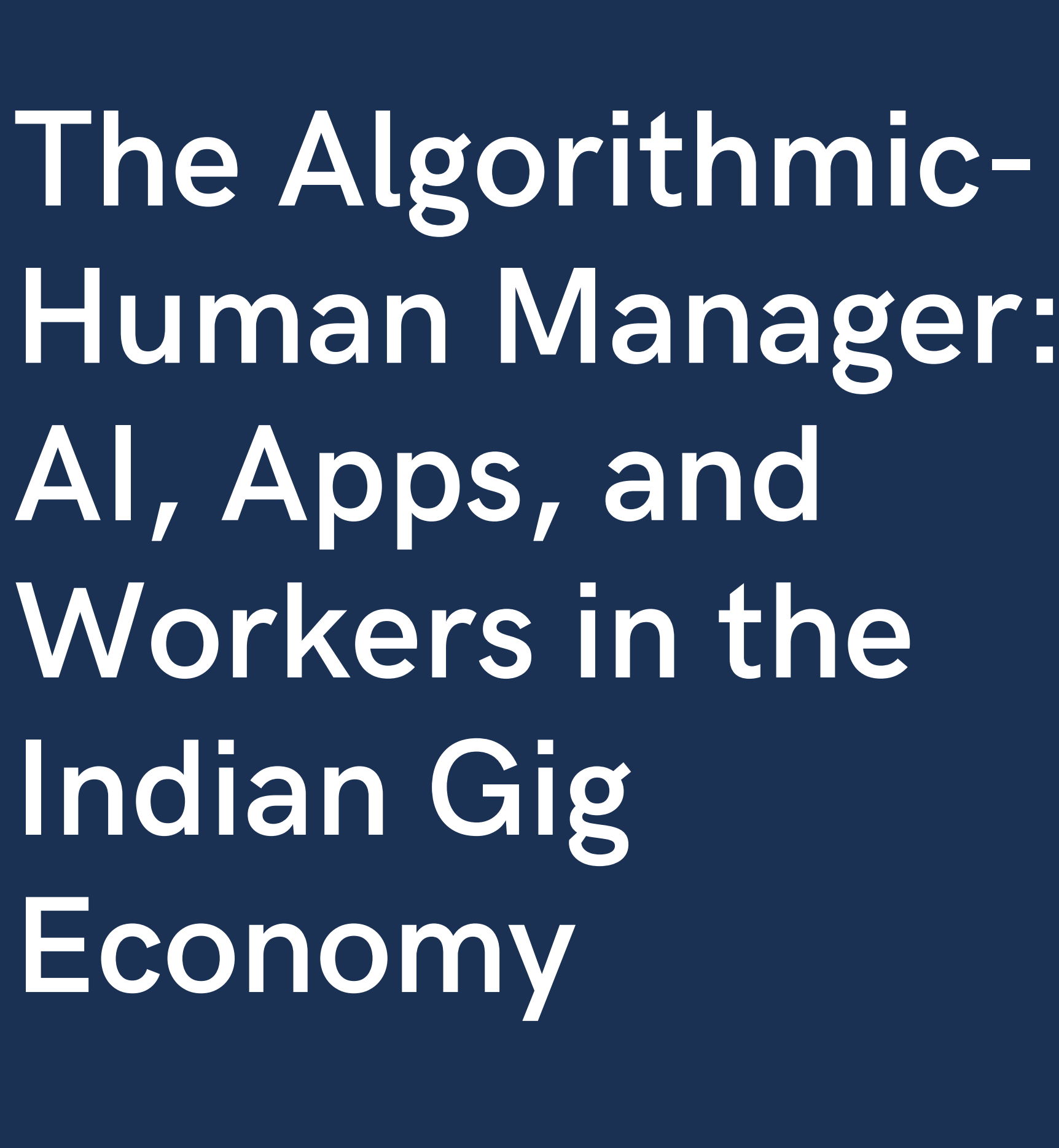

# The Algorithmic-Human Manager: AI, Apps, and Workers in the Indian Gig Economy

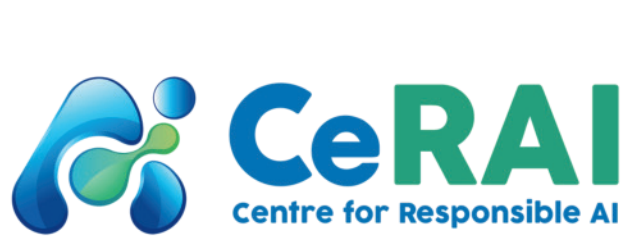

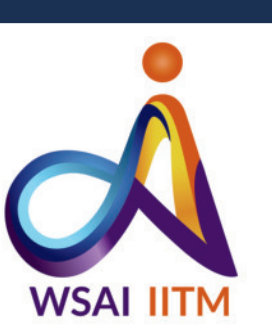

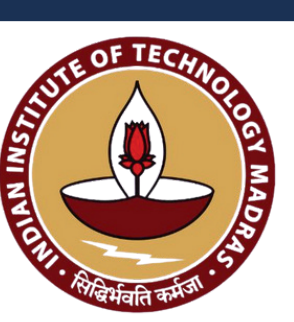

# Authors

Omir Kumar — Policy Analyst, Centre for Responsible AI, IIT Madras (CeRAI)

Krishnan Narayanan — Researcher, CeRAI, IIT Madras; Co-founder and President, itihaasa Research and Digital

# Acknowledgements

The authors express their deepest gratitude to Prof. Balaraman Ravindran, Head, Wadhwani School of Data Science and Artificial Intelligence (WSAI), and Centre for Responsible AI, (CeRAI) IIT Madras, and Prof. Sudarsan Padmanabhan, Professor, Department of Humanities and Social Sciences, IIT Madras & Principal Investigator, CeRAI for their guidance and mentorship throughout the course of this work.

We would also like to acknowledge Bhumika Sharma and Srivatsan S for their diligent research and review assistance.

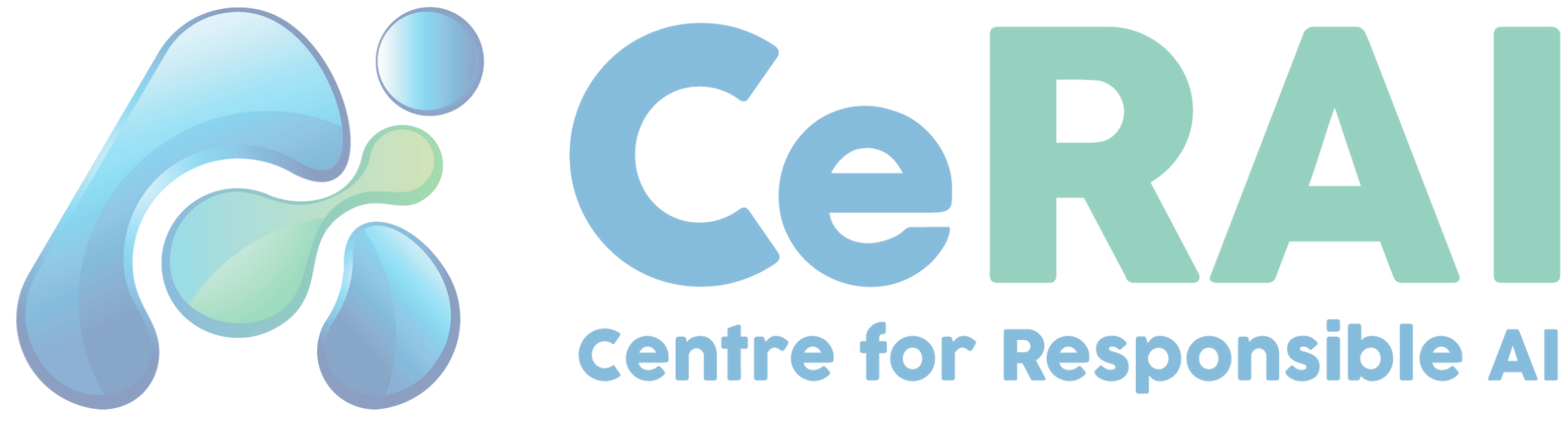

# Executive Summary

The Algorithmic-Human Manager: AI, Apps, and Workers in the Indian Gig Economy, by the **Centre for Responsible AI (CeRAI) at IIT Madras**, examines the profound impact of Artificial Intelligence (AI) and digital technologies on the burgeoning blue-collar gig economy in India. With the sector, projected by NITI Aayog, to employ over 2.35 crore workers by 2029-30, understanding the interplay between technological innovation and worker opportunity/welfare is critical.

The study adopts a social justice framework to analyse how algorithmic management shapes the lives of workers in location-based services like ride-sharing, delivery and home services. Through a mixed-methods approach, including in-depth interviews with **16 gig workers and 21 key stakeholders**, the report uncovers a dual reality:

- While AI and technology creates unprecedented access to work and new efficiencies, it also introduces significant challenges related to fairness, transparency and worker dignity.
- While AI is responsible for the algorithmic management of gig work (allocation, performance management etc.), workers believe that a human is making these decisions.

### What we heard from workers

- **Opportunity – often as fallback:** Platforms expand access to work (e.g., workers moving from unemployment to steady earnings), yet many choose gig work due to limited alternatives rather than preference.
- **Apps as the "manager," but humans behind the curtain:** Workers experience core decisions (order distribution, pay, incentives) through the app but believe human teams ultimately control rules. While their professional lives are governed by efficient algorithms, they neither understand nor trust them. This opacity in task allocation (e.g., favouring longer hours or new joiners), wage determination, and performance ratings fuels a sense of powerlessness, particularly around opaque pay structures and penalties for factors beyond their control (e.g., traffic).

- **Grievances need a human touch:** Chat support is the most-used feature but struggles with complex issues (penalties, deactivations, payouts). Workers want time-bound escalation to a human; chatbot-only pathways are viewed as unfair and alienating.
- **Worker Perceptions of AI:** Most blue-collar workers are unfamiliar with the term "AI", yet describe the perceived effects of AI on their work. Their primary interaction with technology is the platform's app, which they often view more as a tool of control than one of empowerment. They do not believe technology can fully replace them, citing real-world complexities like poor infrastructure.

### What stakeholders told us

- **Digitised recruitment & onboarding**: Firms use targeted ads, data pipelines, and even GenAI to source and process candidates. Workers can onboard themselves with minimal cost and begin earning in hours, a process that once took days. Inclusion features (e.g., speech-to-text) help, but digital literacy barriers, reduced "human touch", and rural intermediaries impose new frictions and risks.
- **Algorithmic Management's Double Edge**: Platforms use technology to streamline recruitment and manage performance at scale, employing tools like selfie verification and location tracking to prevent fraud. However, stakeholders also describe practices of "algorithmic cruelty", such as the perceived manipulation of incentives and deactivations without due process. Practices like "income obfuscation" and digital surveillance create a precarious work environment, undermining the very flexibility that attracts workers.
- **Human-tech collaboration in warehouses/dark stores**: Handhelds, computer vision, and layout optimisation lift throughput; yet rigid targets ignore weather/flooding/heat. Automation helps in harsh conditions but can also hard-code unrealistic expectations, exposing workers to penalties for delays beyond their control.
- **Wages and incentives – visibility with opacity**: Some apps show real-time earnings and surge logic; still, many workers face income obfuscation, unilateral rate changes, and arbitrary penalties that can wipe out weekly income, creating financial fragility and stress.
- **Policy landscape is a work in progress**: The Code on Social Security (2020) recognises gig/platform workers but provisions for gig workers have not been notified yet. Some states (Rajasthan, Karnataka, Jharkhand, Bihar and Telangana) are advancing measures on transparency, grievance redressal, algorithmic explainability and social-security contributions. Implementation, data rights and "human-in-the-loop" standards remain uneven.

### The Path Forward: A Pragmatic Middle Path for an Equitable Gig Economy

The report advocates for a "pragmatic middle path" that fosters innovation while establishing robust digitally-enabled guardrails to protect workers. Key recommendations are targeted at all ecosystem participants:

- **For Platform Companies**: Embrace algorithmic transparency by providing clear explanations of how systems work. Establish human-centric grievance redressal with a human-in-the-loop for high-stakes decisions. Adopt participatory design by involving workers in the creation and auditing of platform features. Embed micro-learning into the worker app.
- **Data Gigs Ecosystem**: Create a "Data Gigs Ecosystem" by forging, through the digital platform, a network of annotation companies, speech/text labeling partners, local training NGOs, and DPI enablers in order to create additional earnings opportunities via digital tasks (such as data labelling and annotation) for gig-workers.
- **For Gig Workers and Civil Society**: Build peer support networks to combat isolation. Expand digital and financial literacy initiatives. Leverage technology and data analysis to aggregate best practices, complaints and evidence patterns at scale.
- **For Governments and Policymakers**: Enable a foundational safety net by developing a Unified Worker Interface (UWI). Promote co-regulation to address "design / dark patterns" in gig work and enact a legal "Right to Explanation and Human Review" for critical automated decisions.
- **Unified Workers Interface (UWI)**: A digital public infrastructure (DPI), akin to UPI/ONDC, for the market of work: portable UWI-ID, discovery / interaction / settlement protocols, neutral governance, and reputation portability across platforms. UWI can operationalise social security (contributions, benefits) by creating a verified, consented ledger of earnings and work history, while restoring data ownership to workers.
- **Public repository of design & dark patterns**: A transparent catalogue of harmful and helpful patterns (e.g., wage concealment, false urgency, unfair deactivation; vs. explainable pay, stable scheduling). This enables self-assessment, public accountability, and targeted remediation by platforms and regulators.

We deliberately use the term Algorithmic-Human Manager to capture India’s lived reality and the direction we advocate: algorithms may allocate and optimise, but accountability, fairness, and dignity demand a human in the loop, someone who can explain, override, and redress.

This hybrid governance reframes the debate from “algorithms versus people” to “algorithms with people”: clear, auditable systems paired with time-bound escalation, worker participation, and due process so efficiency and equity reinforce each other. By adopting this middle path, India can guide the evolution of its gig economy towards a future that is not only economically dynamic but also socially just and equitable for the millions of workers who power it.

# 1. Introduction

## 1.1. AI, Digital Technology and Work

Technological waves - from steam power and the mechanical loom to assembly lines, computing, and now artificial intelligence (AI) - have continually reshaped the workforce and how we work. Their impact can be seen in three main ways: (i) by substituting for tasks through automation, (ii) by creating new tasks, roles, and even sectors, and (iii) by reconfiguring how existing jobs are performed (Autor, Levy, & Murnane, 2003; Bresnahan et al., 2002; Autor, 2015; Acemoglu & Restrepo, 2019). In today's AI wave, software (machine-learning systems for prediction, matching and optimisation) and smart hardware (robots, sensors and autonomous machines) together automate parts of both digital and physical work. The International Labour Organisation estimates that overall 2.3% of employment (or 75 million jobs) is at risk of automation because of high exposure to generative AI technology (Berg, 2024).

Against this backdrop, research converges on the following set of practical questions. First, **job displacement and creation**: how much work will be automated, which occupations stand exposed, and where will new work be created as productivity and demand rise (Acemoglu & Restrepo, 2019; International Labour Organization, 2023). Second, **productivity and innovation**: when and how AI actually lifts output, typically when paired with complementary investment in people, data, and processes (Brynjolfsson, Rock, & Syverson, 2017). Third, **skills and education**: the mix of technical and socio-cognitive capabilities workers now need; and how reskilling can narrow, rather than widen, gaps (Autor, 2015; International Labour Organization, 2023). Fourth, **sectoral differences**: uneven effects across industries as task structures differ (Bresnahan et al., 2002). Fifth, **wages, inequality, and polarisation**: potential gains for some high-skill roles and pressures on routine middle-skill work, with distributional consequences (Autor, 2015). Sixth, **fairness and discrimination**: how to prevent algorithmic bias in hiring, allocation, and evaluation (Barocas & Selbst, 2016). Seventh, l**abour-market dynamism**: the churn, mobility, and career transitions AI may accelerate (Autor, 2015; International Labour Organization, 2023). Eighth, and central to this report, **platform work and algorithmic management**: the use of algorithms to match, route, price, rate, and monitor work on digital platforms, shaping autonomy, earnings, and protections (Kellogg et al., 2020; Rosenblat & Stark, 2016; International Labour Organization, 2021).

## 1.2. Platform-mediated or Gig Work

Platform-mediated work, involving a digital labour platform and digital platform workers, is often used interchangeably with gig work or on-demand work. The term "gig" itself has musical origins, first appearing around 1915 when jazz musicians coined it to refer to individual performances (Woodcock & Graham, 2020). By the 1920s, musicians were regularly going from "gig" to "gig", establishing the concept of project-based work. Some key features of gigs are that they are, in essence, non-permanent work and have variable working hours, involving payment on a task-to-task basis (Woodcock & Graham, 2020).

We produce a definition of the key terms as defined by the International Labour Organization - See Box 1 (International Labour Organization, 2021; International Labour Conference, 2025).

(a) The term "digital labour platform" means a legal person or, where applicable under national law, natural person that, through digital technologies, using automated decision-making systems,

i. organises and/or facilitates work performed by persons for remuneration or payment, for the provision of service, upon request of the recipient or requestor.
ii. regardless of whether that work is performed online or in a specific geographic location.

(b) The term "digital platform worker" means a person employed or engaged to work:
i. for the provision of service organised and/or facilitated by a digital labour platform.
ii. for remuneration or payment.
iii. regardless of their classification of status in employment.

(c) There are two kinds of digital platforms:
i. location-based platforms, where services are provided by individuals in a specific location (food and grocery delivery, cab hailing, etc), and
ii. online platforms, where workers provide their services remotely (freelance writing, graphic design, etc).

**Box 1: Definition of the key terms in platform economy (Source: International Labour Organization)**

Today's digital platforms intermediate access to customers and, via algorithmic management, control key elements of work: task access and allocation, performance evaluation and ratings, pricing and incentives, and payment flows (Kellogg et al., 2020; Rosenblat & Stark, 2016; International Labour Organization, 2021). Studying AI's impact on gig work is critical and an emerging area of research because platform algorithms now determine access to work, directly shaping worker earnings, autonomy, safety, and voice.India seems to offer a unique opportunity to study the dynamics between digital tech and the gig economy.

## 1.3. Gig Economy in India

In the last decade or so, India has undergone a remarkable digital transformation. The State of India's Digital Economy Report, 2024 (Deepak et al., 2024) notes that India ranks as the third most digitalised economy in the world overall, and stands 12th among G20 nations when it comes to digitalisation at the individual user level. Access to the internet, financial inclusion, and digital identity have been the three core pillars of India's digital economy. Projects such as Aadhaar, Jan Dhan Accounts, and UPI (Unified Payments Interface) have been key drivers of India's digital revolution (Ministry of Information & Broadcasting, 2024).

India's digital infrastructure and the COVID-19 pandemic enabled the growth of its gig economy. The pandemic disrupted over 12.2 crore jobs during the April 2020 lockdown, with around 75% of them being small traders and wage-labourers (The Hindu, 2021). During this time, the platform economy emerged as a viable source of income for individuals seeking work opportunities. Services like ride-hailing and home services came to a halt during COVID-19, but other services like food and grocery delivery thrived (International Labour Organization, 2024). A study on Indian food delivery platforms highlighted that such platforms provided a buffer for workers when they faced job loss or income disruptions - 9% of workers responded with job loss as a reason to join the platform, whereas about 32% responded that they were unemployed for 5.4 months before joining the platform (National Council of Applied Economic Research, 2023). Further, a survey by Janpahal highlighted how gig work is the primary source of livelihood for many - 85% of workers reported working more than eight hours a day (Janpahal 2024). Further, the study highlighted that of the 5,000 people interviewed only 2.3% were women (Janpahal 2024). More recent work by Vasudevan (2025) has been instrumental in understanding the lives and experiences of gig workers across South Asia.

NITI Aayog (2022) reported that in 2020-21, around 77 lakh people, making up roughly 1.5% of India's total workforce, were employed in the gig economy. By 2029-30, this number is anticipated to grow to 2.35 crore, with gig workers expected to account for 6.7% of the non-agricultural workforce and 4.1% of the country's overall employment. More recent estimates suggest that this number is expected to be around 6.16 crore by 2047 (V.V. Giri National Labour Institute, 2025).

Estimates by Assocham - Primus Partners (2021) suggest that about 64% of India's gig workforce is in the 24-38 age bracket. The cities of Bangalore, Delhi, Chennai, Pune, and Mumbai had the highest number of gig workers in India (Assocham - Primus Partners, 2021). According to Boston Consulting Group and Michael & Susan Dell Foundation (2021), the Indian gig economy could contribute an additional 1.25% to the nation's GDP and generate as many as 90 million jobs over the long run.

Gig workers are engaged in several industries such as retail trade and sales, transportation, and manufacturing. As per NITI Aayog (2022), in 2019-20 nearly 47% of gig workers were engaged in medium-skill occupations, around 22% in high-skill roles, and close to 31% in low-skill work. The data indicates a gradual reduction in the share of medium-skill jobs, alongside a rising presence of both low-skill and high-skill roles.

Gig work is attractive for workers due to the flexibility (no fixed working hours), ease of entry (in terms of assets, investment, and skill set) and ability to diversify income sources. At the same time, some concerns have emerged such as the absence of minimum wages for workers, lack of social security benefits, algorithmic opaqueness, and long working hours for workers (All India Gig Workers' Union (AIGWU), 2023; Jain, 2023; Sakshi Sadashiv K., 2025). All this has led to several studies and policy proposals in India that seek to address these issues.

## 1.4. Studies and Policy Proposals on Gig Work in India

Studies by NITI Aayog (2022), National Council of Applied Economic Research (2023), and IDinsight (2025) revealed insights about the overall gig economy, including the types of services offered, the number of workers engaged, their demographics, motivations to undertake gig work and earnings.

Further, studies by IT for Change; Centre for Labour Studies, NLSIU (Sen, Chatterjee, Agarwal, & Sharma, 2023) and PAIGAM, University of Pennsylvania (2024) have been instrumental in bringing forward issues affecting gig workers, such as working hours, low earnings, lack of social security, and arbitrary control by algorithms. Some work has been done around the use of algorithms and their impact on gig workers, such as arbitrary deductions from payouts (Madhulika et al., 2025). Fairwork India (2023) has contributed by studying platforms and providing them annual ratings based on their treatment of workers across parameters like fair wages, contracts, and social security.

The Code on Social Security (2020), passed by the Indian Parliament, was the first piece of legislation in the country on gig work. It defines a gig worker as "a person who performs work or participates in a work arrangement and earns from such activities outside of a traditional employer-employee relationship". It has several provisions for gig workers, including social security schemes such as accident insurance and health, maternity benefits, and old-age protection. The central government has also been working with platforms to onboard gig workers on the e-Shram portal to create a national database of all gig workers (Ministry of Labour & Employment, 2025b). This database will be important to identify gig workers and give them access to social security benefits (Ministry of Labour & Employment, 2025a).

Further, some states like Rajasthan, Karnataka, Jharkhand, Bihar and Telangana have proposed their gig work laws. (The Rajasthan Platform Based Gig Workers (Registration and Welfare) Act, 2023; The Karnataka Platform Based Gig Workers (Social Security and Welfare) Bill, 2025; The Draft Telangana Gig and Platform Workers (Registration, Social Security, and Welfare) Bill, 2025; The Draft Jharkhand Platform Based Gig Workers (Registration and Welfare) Bill, 2024; The Bihar Platform Based Gig Workers (Registration, Safety and Welfare), Act, 2025.

Broadly, all these state specific laws seek to — (i) provide social security benefits to gig workers, (ii) put certain obligations on aggregators relating things like grievance redressal, termination of work, algorithms deployed on the platform, and contribution towards social security schemes, and (iii) provide for the registration of aggregators and gig workers.
In this context, and given the increasing use of AI and digital technologies, there is a need to understand their potential impact on blue-collar workers engaged in the platform/gig economy in India.

# 2. Study on the Impact of AI & Digital Technologies on Blue-collar Gig Work in India

## 2.1. Research Organisation

**The Centre for Responsible AI (CeRAI)** at IIT Madras is a leading interdisciplinary research centre located in the Global Majority, dedicated to advancing both the technical and policy dimensions of Responsible AI. CeRAI brings together fundamental, applied, and policy research to advance the development and deployment of AI systems that are ethical, fair, and accountable. CeRAI has conducted this study to understand the **impact of AI and digital technologies on blue-collar gig work in India**.

This study also falls under the research focus of the **Co-Intelligence Network at IIT Madras**, which studies the five futures of a co-intelligence world: learning, planet, development, work and AI. This study comes under the Future of Work.

## 2.2. Scope and Outcomes

This study focuses on the use of AI and digital technologies by **location-based platforms in the Indian gig economy**. This includes services like ridesharing, delivery services, and home services. The study seeks to understand how platform companies are leveraging AI and other digital technologies to transform their various functions. Further, the study unpacks the experiences of workers engaged with such platforms.

There is widespread concern that these technologies may displace or cause harm to low-skilled workers, even as they create new employment opportunities. That is why the study analyses these impacts from the lens of **social justice**, recognising that technological deployment must not only improve efficiency and scale but also **promote fairness, equity, and dignity for workers**.

Our analysis draws on John Rawls's justice as fairness (Rawls, 1971), using the difference principle to test whether AI-enabled arrangements improve the position of the least advantaged among gig workers.

The Study:

- **Maps** the use of AI and digital technologies across the gig economy ecosystem, examining how platforms deploy these tools to automate or augment tasks across the value chain.
- **Assesses** the impact on workers, capturing both the objective effects (earnings, hours, conditions) and workers' own perceptions of these technologies.
- **Analyses** findings through a social-justice perspective, focusing on fairness, inclusion, capability-development and worker well-being.
- **Provides** practical recommendations and guidance for platform companies, workers and policymakers to ensure the responsible use of AI and digital technologies in the gig economy.

## 2.3. Methodology

The study employed a mixed-methods approach combining:

- **Secondary research** to establish the landscape of AI and digital technology deployment in gig work. One of the authors of this study used to work at an Indian gig work platform, and insights from these experiences have informed the study.
- **Primary research** with blue-collar gig workers through consultations and interviews to capture their ground realities and perceptions of algorithmic management.

**o** To capture the lived experiences of gig workers, the study employed semi-structured interviews with blue-collar platform workers across services such as ridesharing, food and grocery delivery, and home services. A guiding questionnaire was used, covering themes such as daily work practices, awareness and perceptions of AI, changes in work due to platforms or automation, fairness and transparency of algorithmic decisions, income and job security, access to training, and workers' aspirations for better protections and conditions.

**o** The interviews with gig workers were conducted in person (with a few on call) in the New Delhi/NCR region and Bengaluru, allowing researchers to interact directly with workers and observe them in their workplaces, including dark stores and other operational settings.

o The semi-structured format allowed for consistency across interviews while leaving room for workers to share their own narratives and examples in depth, including in their local languages (Hindi / Kannada). This approach ensured both comparability of responses and rich qualitative insights into how AI and digital technologies are shaping the realities of gig work in India.

- **Primary research with other stakeholders** including platform providers, industry leaders, civil society actors, and policymakers to obtain a holistic view.
  - For platform companies and industry leaders, discussions focused on the deployment and rationale for AI and digital technologies, measurement of their impact, worker outcomes (both positive and negative), fairness and bias safeguards, and future skill requirements.
  - Civil society stakeholders were consulted on observed worker experiences, adequacy of worker representation, principles for ethical technology deployment, and pathways for collaboration to balance innovation with worker protection.
  - The interviews with these stakeholders were conducted virtually via video conferencing, as participants were based across different parts of India.

# 3. Findings from the Primary Interviews with Gig Workers

In the platform economy, workers' interactions are consistently mediated by multiple technologies. Examining how workers perceive and respond to these technologies is essential for understanding the dynamics of gig work. Accordingly, we spoke to 16 gig workers. The following are key observations.

## 3.1. Gig Work - Technology Creates Opportunities, But Not Always by Choice

Workers recognise that technology has benefited them by creating work opportunities (i.e., gig work) and making access easier.

- One worker (an electrician) highlighted how he was unemployed before joining the platform. He tried to get work opportunities by physically approaching potential customers, asking for work, and distributing his visiting card but in vain. He then joined a gig work platform and is now able to provide for his family and pay off his debt.

While workers acknowledged the benefits of gig work, they also highlighted structural challenges in the labour market, such as unemployment and the shortage of well-paying jobs.

- For many workers, gig work is not the first choice but rather one of the few viable options available.
- One worker (driver) claimed that because of the platform, non-local drivers also get opportunities. And they drive rashly as compared to the locals.

## 3.2. Applications as the Core of Work, But Believe Decisions Are Human-Made

Technologies, particularly the algorithms embedded in applications ("app"), play a central role in shaping how work is organised, including task allocation, pay, and incentives. However, workers feel that their work is being controlled not by the application itself but by humans.

- When asked who makes decisions on order distribution, pay, and incentives, workers mentioned roles such as supervisor, team leader, central team, store manager, and even "a team in Bangalore".
- Many workers opined that while decisions are implemented through the app, they are ultimately decided by humans.

This perception of an unseen, human-led control system, even when decisions are automated, points to a fundamental power imbalance that challenges the principle of 'justice as fairness'. From a Rawlsian (Rawls, 1971) perspective, the opacity of the system creates an inequality where workers, the least advantaged group in this relationship, lack the information and agency to understand or contest decisions that directly impact their livelihoods. The lack of consultation mentioned by one worker further underscores how the system's benefits, such as efficiency, accrue disproportionately to the platform rather than improving the position of the workers themselves.

Workers feel that there is a lack of transparency in how decisions are made, and they do not believe the app is fair in pay and work allocation.

- One worker mentioned that even if companies explained how the app works, workers would not understand the "algorithms" and the "maths" behind it.
- One worker mentioned how the company unilaterally takes decisions and does not consult them.
- Some workers also observed that the app gives more work and higher payouts to those who work longer hours.

During fieldwork in Delhi, we observed a strike by quick-commerce gig workers who switched off their IDs and stopped accepting orders. The protest centered on payouts: workers said their pay had been cut while new hires earned more for the same work. Some whose pay was unchanged joined in solidarity. Finally, the company raised payouts and the strike was called off. Workers raised fairness concerns about the app's selfie verification. Platforms use these checks to prevent ID sharing and confirm uniform compliance, but the system can be unreliable: one worker submitted a selfie while wearing the company T-shirt yet was penalised for not wearing it, and another said his friend sometimes uses his ID without the selfie check catching it.

The platform and workers often differ in their notions of fairness. For instance, the platform may define fairness as assigning tasks within a 5 km radius or recording a trip as a late start even if delayed by only five minutes, whereas workers perceive these situations differently. As some drivers explained:

- Depending on some traffic hotspots in the city, it takes longer to reach the destination to pick up a customer. But there is no leeway provided.
- In Bengaluru, covering a distance of 3–3.5 km by auto typically costs ₹80–90. Earlier, two-wheeler ('bike-taxi') services were available at roughly half that price.
- One driver who worked on a "drivers for hire" platform highlighted how they get deactivated from the platform if a customer complains about an accident, even when the accident is caused by someone else. "These food delivery people drive their two wheelers very rashly".

This disconnect in perceptions of fairness can be viewed through Sen's (1999) capability approach. While platforms offer the flexibility to work, the rigid and unforgiving nature of the algorithms, such as penalising a driver for a five-minute delay in heavy traffic, constrains a worker's real capability to perform their job effectively and without stress. Their autonomy is limited, and their ability to navigate real-world complexities is not accounted for, thus reducing their freedom to lead a life they value while working.

## 3.3. Workers Engage the most with Chat Support; Need for Human Touch

When talking about the application, all workers brought up chat support (grievance redressal). While the application has several features, it seems that workers engage mostly with chat support. This is an important feature for them. One worker remarked: "When we are out for deliveries, it's only the chat support that is there to help us."

Chat support works well for time-sensitive issues that arise during gigs (such as cancelling rides/orders or contacting customers). However, for other issues related to payouts, incentives, penalties, and ID deactivations - workers feel that the system has problems. These include: (a) absence of time-bound resolution of complaints, (b) inadequate resolutions, or (c) no justification provided for certain decisions.

The shift to automated and inadequate grievance redressal systems, such as chatbots that fail to understand complex issues, is a clear example of what Galtung (1969) would term 'structural violence'. This structure prevents workers from meeting their basic need for security and dignity by denying them a meaningful way to resolve critical issues like pay disputes or unfair deactivations. This creates a state of perpetual insecurity, where workers are vulnerable to algorithmic errors or biases without any effective recourse, embedding harm into the operational design of the platform itself.

A recurring issue raised by workers was the option to talk to an executive. Workers mentioned that they raise a ticket and then get a call from the support team, while only some platforms allow direct calls.

- One worker recounted that the platform he previously worked for had integrated a chatbot into its customer support system. He said: "The chatbot does not listen to or understand our issues. It only responds based on the data fed into it."
- Workers recommended that they should be able to call the support team and speak directly to a human resource and there should be time-bound resolution of complaints.

## 3.4. Make Hybrid / Online Training More Engaging

Given that most functions are performed through the application, platforms provide online training to workers covering topics like app usage, grievance redressal, platform guidelines, and code of conduct.

- Almost all workers mentioned that they had watched these training videos while onboarding. Only a few said they had not seen such videos.
- One worker said he had not received any training. When asked if he had seen the training videos, he said, "Yes", and added: "Training should be in person and not online."
- Most workers feel that companies should offer in-person training for workers during onboarding. A few said companies could make their online training content more engaging.
- One worker, when asked to describe the features in the app he doesn't use, mentioned the training videos tab. These companies create content and make it available for workers, but the content was never used.
- Other tabs not seen include the analytics dashboard of the worker's operational data - he had never seen the tabs providing various reports on parameters like acceptance rates and on-time gig completions.

On the effectiveness of online training videos, most workers said they are not effective for first-timers or digitally illiterate workers. These workers often struggle to understand the workflow and functioning of the application and rely more on experienced workers for support during their initial weeks.

## 3.5. AI: Workers Are Unaware, But Have Opinions

We asked workers whether they were familiar with AI. Most said no, though some had heard of it without knowing if companies actually use it.

- One worker remarked - "I have heard of AI but the company does not use it."
- One worker suggested that companies should use AI to optimise order allocation. He explained that he often gets orders in the opposite direction of where he is going and would prefer AI to allocate orders along his route or nearby. He remarked,"All future jobs will be AI related so people who do an AI course should be secured."
- One worker mentioned "Tesla and automatic driving" as an example of AI. But when asked if he has ever used AI, he said, "No." He also felt there was no AI in the app he used.

When asked if technology or AI could replace their work, most workers said no. Their reasons included:

- Machines cannot handle infrastructural issues such as potholes or electric wires. However, one worker said that drones will be able to effectively navigate such issues as it will be controlled by a human.
- Customers want doorstep delivery, and machines cannot climb stairs.
- If machines replace workers, there may be large-scale protests opposing automation.

Some workers mentioned having seen videos of drones making food or grocery deliveries but were unsure if those were real or fake.

- Talking about humans getting replaced with machines, one worker remarked, "All technology and innovation is done by humans so no matter how much technology advances, humans will always play a critical role."

During our discussions about automation in dark stores, a store-level executive said that while robots could replace human workers, he would not prefer it.

- He explained that if machines break down, it may take two to three days to fix them, during which store operations would stop.
- He added that unlike humans, machines cannot be pushed to work faster or more efficiently.

We also heard that automation in dark stores is difficult due to their varying sizes across the country. It will be difficult to standardise the size of machines, the layout and processes across stores. On the same issue of automation, a worker (who used to work in a dark store) mentioned that machines are not as fast as humans in picking and packing items. Machines aren't instinctive; they won't be able to think on the spot.

- He gave an example - if a machine does not find an item in its designated spot, it won't know what to do whereas a human worker will look around or ask someone and get the item.

# 4. Findings from the Primary Interviews with Other Stakeholders

The proliferation of gig work platforms and AI has profoundly reshaped India's labour market, particularly in the blue collar workforce. This technological wave presents a dual reality: on one hand, it has unlocked unprecedented efficiencies and created new avenues for employment; on the other, it has introduced complex challenges related to worker rights and welfare. For gig work platforms, the primary drivers for adopting these technologies are clearly articulated as cost reduction and increased efficiency. For workers, however, the often-subtle distinction between digital technology and AI is irrelevant; it is simply the new reality of their work, mediated entirely through a smartphone app.

We spoke to 21 platform executives, business leaders and worker advocates from the civil society. Insights gathered from them highlight the following themes: i) the digitisation of recruitment and onboarding, ii) the rise of algorithmic management - work allocation, performance management, wage determination, iii) the state of grievance redressal and worker safety, and iv) the evolving regulatory framework designed to govern this new world of work.

## 4.1. The Digitisation of Recruitment and Onboarding

Conversations with stakeholders revealed that recruiting blue-collar workers continues to be a significant challenge due to three main factors: i) high attrition rates; ii) limited channels for companies and workers to connect; and iii) the inability of third-party hiring agencies to consistently identify suitable candidates.

The business leaders noted that technology is increasingly helping companies address these challenges. Platforms often outsource recruitment while also encouraging workers to self-onboard through direct advertisement (see Figures 1 & 2) or refer peers.

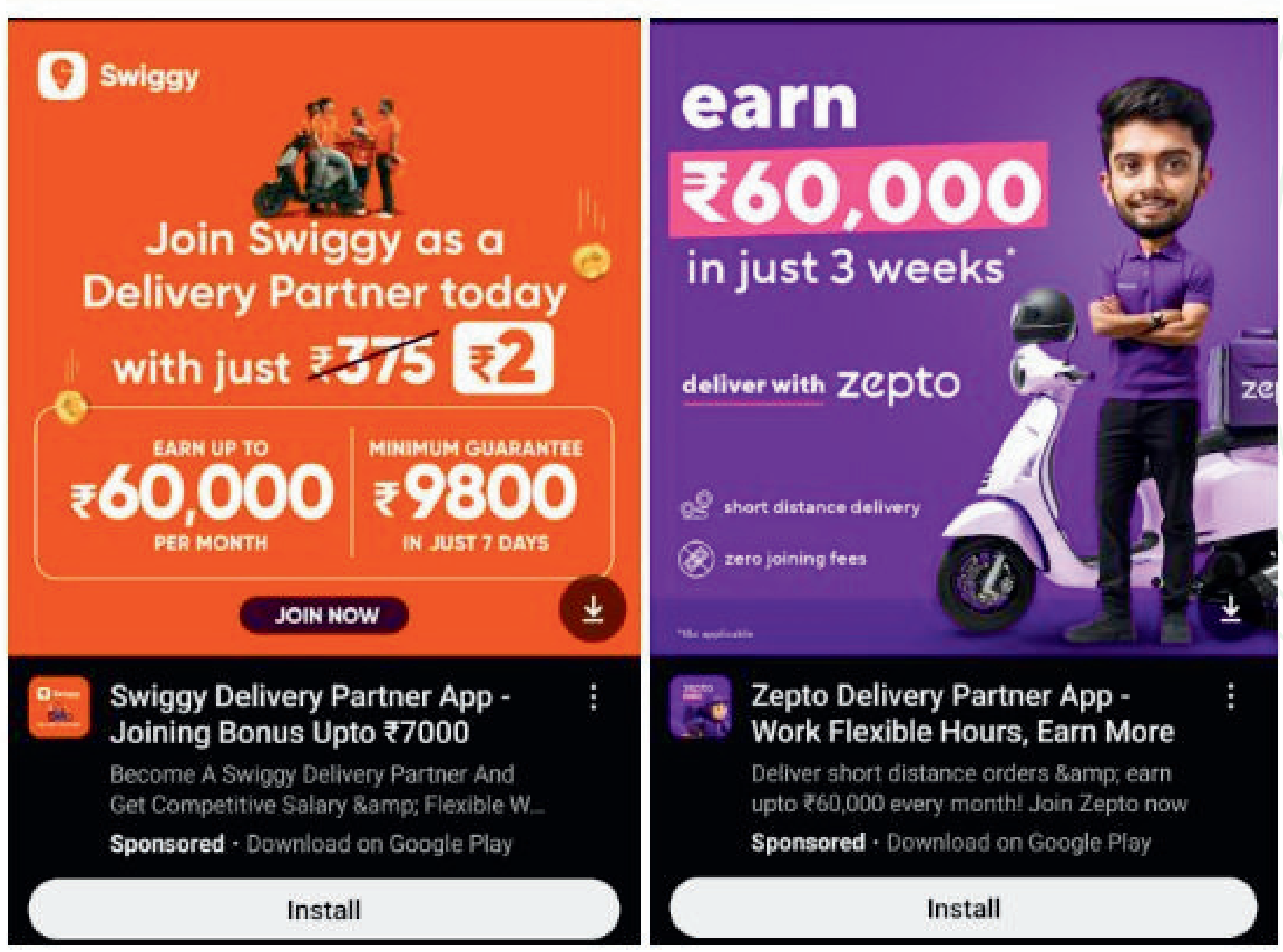

**Figure 1 & 2: Advertisements on YouTube for gig work**

## Tech-Driven Recruitment Improves Access

### Onboarding

Workers today can onboard themselves onto digital platforms with ease and at minimal upfront cost, making gig work highly accessible. Typical requirements include: minimum age eligibility, a smartphone meeting basic specifications, a two-wheeler or four-wheeler compliant with legal and safety norms (for mobility or delivery work), and valid government-issued documents (e.g., driving licence, registration and insurance certificates, proof of address, PAN card, and bank account details). Verification of documents and payment of onboarding fees (to cover document verification and starter kits such as delivery bags or uniforms) are generally completed online. Most platforms also deliver training digitally, covering topics such as how to use the application, platform guidelines, and grievance redressal mechanisms.

### Recruitment

Traditional hiring processes, often slow and localised, are now being transformed by technology-driven systems, involving data analytics and generative AI (GenAI). One of the interviewees stated that a company was using targeted social media advertising to reach potential candidates, collect their details, and even generate resumes using GenAI. AI tools are also employed to tailor CVs according to job requirements.

For platform companies facing high turnover, digital recruitment addresses operational challenges by scaling hiring and reducing onboarding times, from several days to just a few hours. As one interviewee observed, workers are eager to start earning immediately, and accelerated onboarding meets this urgent need. A key benefit of tech-enabled recruitment is that workers don’t need to miss work and lose wages to attend in-person interviews and other processes. Technology is also being leveraged to promote inclusion. Examples include speech-to-text systems to assist workers with low literacy and video-based job descriptions to improve comprehension.

Even with these advancements, technology remains more effective at evaluating objective criteria, such as whether a worker has a bank account or government ID, than subjective qualities like motivation, attitude, or reliability. To address this, some companies are experimenting with AI-driven interviews that assess the candidate’s behaviour and generate candidate scores. Though the early results were promising, questions about accuracy remain to be addressed. However, for every gap it bridges, technology can create new barriers. Low digital literacy among workers leads to high drop-off rates when navigating complex HR processes/ platforms, especially as it forces workers to learn new systems.

Further, while technology increases efficiency, it comes at the cost of “human touch”. The absence of in-person interactions makes it harder to verify intangible qualities such as intent, communication skills, or long-term reliability.

This gap in “soft verification” contributes to the very attrition problem platforms are trying to solve. Another key challenge in gig economy recruitment in India is the role of intermediaries, particularly in sourcing workers from rural areas. Insights from stakeholders indicate that these middlemen often function with little oversight, charging unaccounted fees or misrepresenting the nature of the job. Technology alone cannot solve these deep-rooted practices without coordinated efforts in digital literacy and trust-building.

## 4.2. The Rise of Algorithmic Management – Work Allocation, Performance Management, Wage Determination

Today, all digital platforms — regardless of which sector they operate in (home services, food delivery, ride sharing, etc) — leverage tech-based solutions to carry out their operations at scale, with AI and data science models being embedded into core operational processes. In gig work, algorithmic management is typically defined by five key features: (a) continuous monitoring of workers' activities, (b) ongoing assessment of their performance, (c) automatic implementation of decisions, (d) interactions with a system instead of a human, and (e) low transparency (Mohlmann & Zalmanson, 2017). This "algorithmic boss" oversees and manages key employer functions — from recruitment to daily performance management (Adams-Prassl, 2019).

**Work Allocation**

Work allocation, i.e., how gig workers are allocated tasks on the platform, and the wages of workers are driven by algorithms. Such algorithmic management has democratised access to work, creating a hyper-efficient marketplace. Underlying these systems are business rules that seek to maintain supply-demand equilibrium. The goal is to deploy the optimal number of workers based on real-time and historical conditions through complex geospatial and predictive analytics.

During peak hours or festive seasons, companies offer targeted incentives to encourage workers to log in and meet increased demand. Some platforms even use algorithms as a safety feature, as noted by one stakeholder, automatically logging drivers off after a set number of hours to prevent fatigue. This demonstrates technology's capacity to enhance worker well-being when designed with that intent.

On the other hand, this same system is widely perceived as an opaque and unforgiving digital taskmaster, capable of what some call "algorithmic cruelty". One practice cited was the slowing down of order allocation just as a worker approaches an incentive threshold, a tactic that workers perceive as deliberately preventing them from reaching their goals. One interviewee highlighted that workers frequently raise concerns about newly onboarded workers receiving more orders and better pay compared to experienced workers on the platform.

This practice of algorithmic cruelty directly contravenes the Rawlsian (Rawls,1971) difference principle, which states that inequalities are only just if they benefit the least advantaged. Here, the algorithmic inequality i.e., the platform's superior knowledge and control, is exploited in ways that affect workers from achieving their goals, deepening their financial precarity instead of improving their conditions.

The use of data by these platforms goes beyond worker management to include strategic decisions such as store placement. While data-driven models typically guide site selection, in practice, some dark stores are established for other reasons. For instance, one interviewee noted that a quick commerce company set up dark stores on the outskirts of Delhi reportedly to meet company targets.

**Performance Management**

The performance of workers on the platform also contributes to this algorithmic decision-making process. Performance parameters include login-logout consistency, customer ratings, complaints (especially delays and delivery-related issues), order acceptance rate, and travel efficiency.

Platforms use computer vision and location tracking to detect and prevent fraud while ensuring compliance and safety. For example, some workers may share their login credentials with others to work on their behalf. To counter this, platforms conduct periodic selfie checks to verify that the registered worker is the one completing tasks, and to confirm that the mandated uniform is being worn. Location tracking further supports monitoring by validating worker presence at pick-up and delivery points, optimising routes, and tracking task completion. In ride-hailing, GPS data is also used to monitor speed compliance and enhance passenger safety.

Conversely, practices such as continuous location tracking (even when the worker’s app is turned off, and without the consent of the workers), accessing the device activity of workers to see what other gig work apps they are using, and recording calls made through the platform can create a sense of excessive digital surveillance for workers. The granular control is further exemplified by automated performance management systems that can trigger termination after a set number of infractions, with no clear process for appeal or human inquiry.

One interviewee noted that workers' IDs are sometimes deactivated, occasionally even permanently, without any explanation. They added that in some cases, platforms have deactivated IDs after workers raised concerns or participated in protests about their working conditions.

The use of constant surveillance and the threat of deactivation without a fair process severely limits what Sen (1999) calls worker 'capability'. While workers technically have the freedom to choose their hours, this autonomy is undermined by a system of control that limits their ability to contest unfair ratings or exercise their voice without fear of reprisal. This lack of genuine agency and the constant monitoring restricts their freedom to function effectively and with dignity in their work environment.

**Human-Technology Collaboration in Warehouses**

In the e-commerce and quick commerce industry, warehouses form the backbone of operations. They act as collection centres where goods from retailers, farmers, and wholesalers are stored before being dispatched. Quick commerce typically relies on two layers of warehousing: the mother warehouse and the dark store. Dark stores are mini-warehouses located close to customers, enabling ultra-fast deliveries in just a few minutes (see Figure 3).

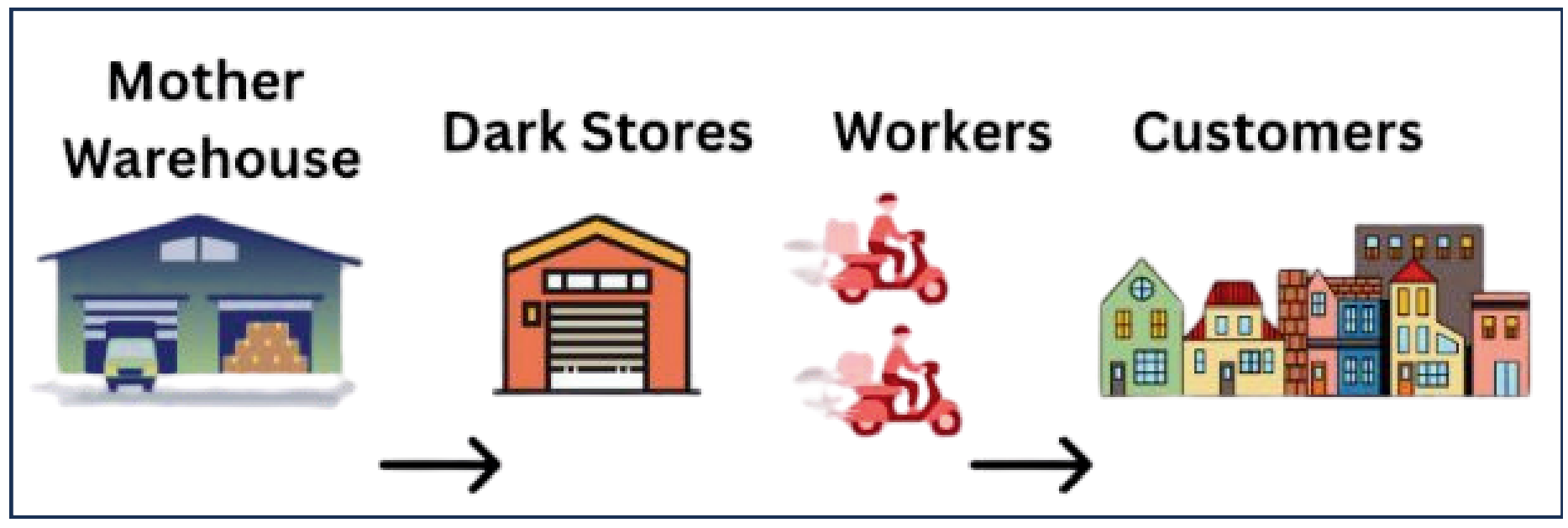

**Figure 3: Flow of goods in quick commerce**

Both warehouses and dark stores are powered by a mix of human effort and technology. Companies use data-driven models to predict demand, optimise delivery routes, and design store layouts. For instance, they place high-frequency items near dispatch counters for quicker access. Workers in dark stores use handheld devices that guide them to the right shelves for picking and packing. Since order picking is the most labour-intensive and costly part of warehouse operations, reducing travel time is a key focus, especially in quick commerce, where speed is critical (Pan & Wu, 2012). These devices also allow companies to track the time workers spend on each task.

Beyond handheld devices, warehouses increasingly use tools like conveyor belts and computer vision systems. One stakeholder executive explained how computer vision is applied for defect detection: orders prone to breakage, such as eggs, are scanned before packing. The system flags complaints or potential issues, which are then validated by cameras. However, the stakeholder noted that the technology is less effective for products like fruits and vegetables, where defects are harder to detect.

An interviewee talking in the context of e-commerce highlighted that automation, ranging from robotic arms to conveyor belts, has been introduced to support workers, particularly in challenging conditions such as extreme heat. Yet, these technologies also impose a high degree of control. Task allocation, targets, monitoring, and even penalties are dictated by automated systems. Also, these systems often fail to adapt to real-world conditions. For example, the time given to complete tasks does not change during periods of intense heat or heavy rain.The interviewee described a system at a company that reportedly continued to enforce performance targets even when the warehouse was flooded, illustrating a profound and dangerous disconnect from human realities. As a result, workers face the risk of getting penalised for delays that are outside of their control.

**Wages and Incentives Determination**

Compensation in the gig economy is a major point of contention, characterised by a mix of real-time visibility and opacity. On the positive side, some platforms provide workers a real-time view of their earnings. This is to empower them by showing exactly how much they have earned at any given moment, including incentives during demand surges or in adverse weather conditions.

One stakeholder remarked that the competitive nature of the market itself enforces a degree of transparency – if a platform does not pay fairly or clearly, workers, who often work on multiple platforms, will simply leave for a better offer. In addition to monetary benefits, companies provide perks such as accident and medical insurance, along with skill development programmes for workers and their children.

Conversely, many interviewees opined that workers experience a system where the breakdown of their wages is difficult, if not impossible, to understand, and where pay structures are determined unilaterally by the platform. The use of complex models and UI/UX interface can serve to "income obfuscation", making it hard for workers to verify their pay or understand why it fluctuates. Rate cards can change without notice, and earnings are often dependent on a variety of factors, including performance of workers, the type of vehicle used and fluctuating surge pricing algorithms.

Furthermore, workers can be subjected to severe and arbitrary penalties that can cripple their income. In one stark example shared by an interviewee, a food delivery worker was penalised Rs. 3,200 following a single customer complaint, a sum that could represent a significant portion of their weekly income.

A survey on food delivery workers and cab drivers in select Indian cities found that 68% cab drivers faced unexplained and arbitrary deductions through company algorithms (PAIGAM, University of Pennsylvania, 2024). This lack of clarity, coupled with the risk of substantial and sudden deductions, creates a precarious financial existence for many, undermining the very stability that gig work promises to provide.

The combination of income obfuscation and arbitrary penalties is another form of structural violence, where the system itself inflicts financial and psychological harm. Just as the inadequate grievance systems create procedural insecurity, the opacity of wage calculations prevents workers from meeting their basic needs reliably. This structural vulnerability is compounded by the legal grey area around liability, leaving workers to bear the risks of the system without adequate protection or a safety net.

## 4.3. Grievance Redressal and Worker Safety

Grievance redressal is a critical part of platform operations, affecting both customers and workers. As our interviewees noted, post-COVID-19, platforms are increasingly relying on chatbots for handling grievances. For workers, this represents a significant shift. Previously, they could approach local company representatives for support, and on-ground staff often visited them to resolve issues.

After COVID, however, the number of such personnel was sharply reduced, leaving chatbots as the primary channel for complaints. As one stakeholder noted, the rise of technology in grievance redressal has come at the cost of human interaction, weakening the direct support networks workers once relied upon.

The shift to technology-mediated work has left many workers feeling voiceless and disempowered. Some of our interviewees remarked that when a decision is made against a worker, there is "no scope for natural justice" or a formal inquiry process. Communication channels that do exist, such as WhatsApp groups, are often controlled by management, and as one interviewee pointed out, a worker can be removed for raising a complaint, effectively silencing them and cutting them off from their peers. The increasing reliance on chatbots for grievance redressal creates an impenetrable wall for workers facing complex issues like unfair pay deductions or arbitrary account suspension.

On the positive side, stakeholders mentioned that technology has also been used to support worker welfare – for example, the automated log-off feature for fatigued drivers, which acts as an important safety measure. Other measures for worker safety include an SOS button on the app for any emergency support and dedicated women-only helpline for safety. However, broader questions around welfare remain unresolved, particularly regarding liability in the event of an accident.

As one respondent put it: Who is responsible – the platform, the worker, or someone else? This legal grey area continues to leave workers in a vulnerable state. Many interviewees stressed the need for a grievance redressal system with multiple levels, where complex issues can be escalated to a human agent instead of being handled only by chatbots. They also called for regular social dialogue, so that workers have a formal space to raise concerns and take part in decisions that affect their working conditions.

## 4.4. Evolving Regulatory Framework

A significant mismatch exists between the nature of gig work in the 21st-century and the 20th-century labour law and social security. The current regulatory landscape may be fragmented and inadequate to address the unique challenges of the platform economy. As a stakeholder pointed out, existing Ministry of Transport guidelines are tailored for ride-hailing and offer little protection for the vast and growing food and grocery delivery workforce. While most respondents acknowledged that The Code on Social Security (2020) is a positive step in recognising gig and platform workers, it has not yet been implemented and only limited provisions have been notified (AZB & Partners, 2023).

State governments have begun introducing policy measures to address concerns around algorithmic management. New regulations are primarily focused on mandating transparency and explainability in how these algorithms function, especially in critical areas like task allocation, wage determination, deductions, and ID deactivations.

Karnataka recently enacted a gig work law requiring platforms to explain, in clear and accessible language (Kannada, English, or any Eighth Schedule language), how workers can access information about automated systems that affect their work. Beyond transparency, a key regulatory push is for the right to a human review of automated decisions. The International Labour Organization (2024, 2025), for instance, has resolved that workers should have the right to request human intervention in high-stakes cases such as payment refusal, suspension, or termination of their accounts.

Another critical tech-related area for regulation involves the protection of worker data rights. Stakeholders have raised significant concerns about the scope of data collection, with instances of platforms tracking workers' locations while they are off-duty or even accessing device activity to monitor the use of competitor apps, often without clear consent.

# 5. Discussions and Recommendations

Navigating the future of the gig economy requires a delicate and strategic balance. The fledgling industry is projected to expand to 2.35 crore workers by 2029-30 (NITI Aayog, 2022) and has the potential to add 1.25% to India's GDP (Boston Consulting Group & Michael & Susan Dell Foundation, 2021). A prudent approach is required, as overly stringent policies risk stifling the gig economy's vital contributions to growth and employment. However, failing to address the fundamental structural challenges confronting its workforce would be a significant oversight that could undermine the sector's sustainable development and worker rights. The path forward demands a concerted effort from all stakeholders in building an ecosystem that harnesses the power of innovation while establishing robust regulatory guardrails, transparent practices, and comprehensive social protections for its most vital asset: its gig workers.

We have chosen the term "Algorithmic-Human Manager" with intent: it names the lived reality of India’s gig workers and the pragmatic future we advocate. Work may be routed by code, but workers experience decisions as human, and they want that human touch when stakes are high. The phrase reframes the debate from “algorithms versus people” to "algorithms with people": transparent models that explain pay and tasking, coupled with time-bound human review; chatbot triage, backed by accountable escalation; participatory design with workers; and DPI-style rails to port identity, reputation, and benefits. It proposes a middle path to governance and regulation where human judgment, dignity, and due process steer automated systems toward fairness. If India builds for this hybrid manager, the gig economy can be efficient and just, creating opportunity and welfare.

## 5.1. Recommendations for Platform Companies

- **Data gigs Ecosystem:** Orchestrate a coalition of partners - data-annotation firms, speech/text labeling vendors, local training NGOs, and DPI enablers - to open a new pipeline of “data gigs” alongside delivery/ride tasks. *See 5.4 Data Gigs Ecosystem*
- **Adopt a Participatory Design Approach:** Regularly and systematically engage with workers in the design and functioning of the platform's applications and algorithms. Offering monetary or non-monetary incentives can encourage participation. This not only empowers workers but also benefits platforms through potentially higher worker retention.

- **Conduct Regular, Collaborative Risk Assessments:** Proactively conduct internal audits on the impact of algorithms on workers. Platforms should work with academic labs to conduct these assessments, as currently, policy is informed by limited studies and anecdotal evidence. Publishing best practices based on these findings can help inform other industry players.
- **Embrace Algorithmic Transparency and Explainability:** Platforms should provide workers with clear, simple, and multilingual explanations of how algorithms affect their work allocation, performance ratings, and wages.
- **Establish a Fair and Human-centric Grievance Redressal System:** Implement a multi-level grievance redressal system where issues can be escalated from chatbots to human agents. Ensure a human-in-the-loop for all high-stakes decisions, such as account deactivations, to guarantee due process and fairness.
- **Embed micro-learning into the worker app**: Treat capability-building like a product using the "learning-in-the-flow" model: daily micro-skilling (context-aware nudges, 5-minute modules, vernacular voice bots, and micro-credentials), adaptive paths, and social learning, so learning directly improves earnings and safety outcomes.

## 5.2. Recommendations for Gig Workers and Civil Society

- **Build Peer Support Networks:** Formal and informal worker groups can serve as crucial channels for sharing information, offering mutual support, and reducing the isolation caused by the shift away from human interaction in grievance redressal.
- **Leverage Technology for Collective Action:** As suggested by experts, worker organisations can use AI and data analysis to process worker complaints at scale, providing data-driven evidence to challenge unfair platform practices.
- **Promote Digital and Financial Literacy:** Enhance programs that help workers understand the digital platforms they use, their rights, and how to manage fluctuating incomes. This is especially important as low digital literacy can be a significant barrier for workers.
- **Conduct Action-Oriented Research:** Work with research institutions to study how responsible AI principles shape platform practices and affect the lived experiences of workers, with the aim of guiding the development of fair, evidence-based platforms and policies.

## 5.3. Recommendations for Governments and Policymakers

The rapid expansion of the platform economy presents a complex challenge for governments worldwide: how to foster innovation and economic growth while ensuring fair and dignified conditions for the millions of workers who power it. The debate often polarizes between two distinct philosophies - i) a "light-touch" approach, champions flexibility and industry-led solutions, arguing that premature or heavy-handed regulation could stifle a nascent and dynamic sector; and ii) a "regulation-heavy" approach, prioritises social justice, advocating for direct government intervention to protect a workforce it views as vulnerable to precariousness and algorithmic control. While both pathways offer valid perspectives, we believe that India must adopt a pragmatic middle path.

- **Enable a Foundational Safety Net via Digital Public Infrastructure**: The government's primary intervention should be activating the Code on Social Security (2020). To ensure efficient and transparent delivery of these benefits, the government could initiate the development of a Unified Worker Interface (UWI). This open protocol would act as digital public infrastructure, similar to UPI for financial payments. UWI can help to streamline the administration of social security contributions and payouts by creating a single, reliable record of worker earnings across all platforms. It also provides workers with ownership of their data. *See 5.5 Unified Workers Interface (UWI).*
- **Promote an Industry-Led Co-Regulatory Framework:** The government can act as a facilitator, encouraging platforms to form an industry association. This body, in a tripartite arrangement with worker representatives and government oversight, would be tasked with developing a "Code of Conduct". This code would establish best practices on fair pay principles, data ethics, and transparent grievance redressal, allowing the industry to regulate itself proactively.
- **Establish a Public Repository of Dark/ Design Patterns for Gig Work:** The government, in partnership with academic institutions, can create and maintain a public repository of known dark patterns and design patterns for workers. By establishing such a repository, platforms would then be encouraged to self-evaluate their systems against this repository and voluntarily report on their mitigation efforts. *Refer to 5.6 - Design/Dark Patterns for Gig Work*.

- **Enact a "Right to Explanation and Human Review":** Mandate that platforms provide clear, simple, and multilingual explanations for any high-stakes automated decision, particularly account deactivations and significant reductions in earnings. This must be coupled with a mandatory "human-in-the-loop" for reviewing and appealing these decisions.
- **Launch a Government-Backed "Fair Work" Rating System:** The government, in collaboration with an Indian academic institution, could manage a public-facing rating system. This system would evaluate platforms against clear metrics derived from the Code of Practice (e.g., average hourly earnings, dispute resolution time, worker satisfaction). Platforms with higher ratings could be promoted on government websites or receive some benefits like tax benefits, empowering consumers and workers to make informed choices.
- **Utilise Regulatory Sandboxes for Future Challenges:** For emerging issues like algorithmic wage-setting and dynamic pricing, governments should promote the use of regulatory sandboxes. Platforms should be invited to test new, more transparent models in a controlled environment, allowing the government to gather real-world data before codifying new rules into the Code of Practice.

## 5.4. Data Gigs Ecosystem

We analyse the opportunity space using the "Co-Intelligence" framework of Ramaswamy and Narayanan (2025). They propose a PIEX lens to visualise new value-creating opportunities - i.e., Platforms of Interactive Engagements driving Impacts in Ecosystems of life-eXperiences. Such a lens helps visualise how value is co-created through Human-AI interactive engagements across organizational ecosystems, focusing on both the process and outcomes of engagement to generate sustainable wellbeing impacts (SWIs).

Platform companies can extend their platform by designing a new set of interactive engagements that orchestrates data-work at scale, such as annotation, transcription, data cleaning, verification, through a coalition of partners. They can map co-intelligence flows between workers (who operate as creative-experiencers - the more they participate, the more they gain), platform algorithms, partner firms, and digital public infrastructures. They should treat the marketplace, quality layer, and learning layer as Shared Digitalized Infrastructure (SDI) capabilities, reusable across task families and partners; and plug-in new domains without rebuilding the stack.

The entire data-gigs ecosystem is optimised for sustainable wellbeing impacts (SWIs) such as income stability, skill mobility, and safety, rather than just output volumes or process efficiencies. A recent example of a data-gigs service (although not of an open ecosystem as we have envisaged here) is from Uber. Uber AI Solutions provides enterprise-grade tools for data annotation, testing, and localization, and as part of this company service, Uber has piloted a program in India where drivers can earn extra income by doing digital tasks within the app (Moneycontrol, 2025).

**Ecosystem Partners:**

- **Platform company** manages task marketplace, matching, quality loop, payments, worker app.
- **Data partners** (annotation/cleaning vendors, tech firms) manage task supply, ontologies, QA rubrics.
- **Plural sector partners** (NGOs, SHGs, skills missions) handle the first-mile recruitment, and provide practical support like digital literacy training.
- **Public sector**: DPI rails (ID, payments), procurement of public-interest data tasks (health, agri, mobility).
- **Knowledge environment** is a lightweight co-intelligence architecture to record ground truth, collect in-context worker feedback, and turn both into bite-size learning assets, such as tips and rubrics, that surface while tasks are being done.

**Data Gigs Progression Ladder:**

**Tier 1 - Foundation ("Collect & Tag-Light")**: Typical gigs: simple, mobile-friendly contributions such as photo and video capture in public spaces or shops, geotag/address checks, basic image/audio tagging, short speech snippets in local languages, OCR clean-ups and yes/no quality flags. These are designed as low-friction interactive engagements that surface in the flow of a worker's day.

**Tier 2 - Skilled Annotation & QA**: Typical gigs: detailed labeling, text intent/sentiment classification, taxonomy mapping, speech segmentation, model-comparison triage ("which output is better and why?"), defect discovery, and scenario edge-case tagging. These tasks help build analytical skills and accuracy habits.

**Tier 3 - Safety & Workflow Leads**: Typical gigs - policy-guided red-flag review, scenario validation, dataset spot-audits for bias/coverage, prompt/response evaluations against documented rubrics, taskset calibration, mentoring cohorts, rubric refinement with data partners. These roles coordinate people-and-AI, keep standards coherent, and close the learning loop.

**Role of the Policy Maker**:
The government / IndianAI mission can seed a "Data-Gigs India" or a "Data-for-Public-Good" public-private program. They can publish open procurement lots for public-sector annotation (e.g., local language, health, agriculture), set quality/ethics standards, and route tasks via interoperable rails (DPI/ONDC) so multiple platforms can offer them. They can also fund community-based AI-literacy and micro-credentialing programs so blue-collar workers can enter, learn, and advance within data work. They may also publish rubric/ethics templates (e.g., to disclose task provenance, pay rates, model-risk safeguards, and data-protection by design) and require SWI reporting (e.g., earnings stability, tier progression, grievance resolution time) as procurement conditions.

## 5.5. Unified Workers Interface (UWI)

In the last decade, India's Unified Payments Interface (UPI) has fundamentally rewired the nation's financial landscape. By creating a shared digital public infrastructure - a common language for value exchange - it has enabled a vibrant ecosystem of applications, fostering unprecedented innovation and financial inclusion. Similarly, Open Network for Digital Commerce (ONDC) is an initiative that aims at fostering a decentralised and open e-commerce network for exchange of goods and services.

What if we could apply this same revolutionary principle not just to money or commerce, but to the very concept of work itself? This is the premise of the Unified Workers Interface (UWI): a protocol designed to create a single, interoperable, and liquid market for human talent in India. UWI is not an application or a job board; it is a foundational set of rules that would allow any worker to be discovered by any employer, on any platform, creating a system that is more efficient, transparent, and equitable for all.

A critical challenge in the modern economy is the classification of gig workers, who are often labelled as "partners" by the platform companies. UWI is uniquely positioned to address this issue not by enforcing a legal definition, but by making the distinction irrelevant from a practical standpoint. The protocol is agnostic to legal labels; it simply facilitates a verifiable agreement between the provider of a "Work Packet" and the holder of a "UWI-ID."

It allows for every completed task and its corresponding feedback to be appended to the worker's permanent, portable UWI-ID. This process formalises their work history by creating a consolidated digital record of their skills, reliability, and income. This verifiable history becomes the key to empowerment, giving workers the tangible benefits often associated with formal employment, such as access to loans, insurance, and social security, without needing the official title.

**The Core Constituents of a Unified Workers Interface**

To function effectively, UWI would need to be built on a set of robust, standardised components, much like the protocols that govern the internet. These would include:

1. **The Unified Worker Identity (UWI-ID)**: The cornerstone of the system, this would be a unique, portable, and verifiable digital identity for every worker (e.g., name@uwi). Linked to a foundational national ID for Know Your Worker (KYW) verification, this ID would act as a universal key, allowing a worker to access opportunities across the entire network. Crucially, the worker would retain full control over their data, managing privacy and deciding what information is shared at each stage of an engagement.
2. **A Multi-Layered Protocol Stack**: UWI would operate through a series of interoperable protocols:
    - **Discovery Protocol**: A standardised method for employers to broadcast "Work Packets" (structured job descriptions) and for workers to signal their real-time availability and skills. This would create a universal search function for the entire labour market.
    - **Interaction Protocol**: Standardised formats for every step of the hiring process, from initial contact and interview requests to digital contracts and non-disclosure agreements.
    - **Reputation Protocol**: A universal system for giving and receiving feedback. This portable reputation, linked to the UWI-ID, would aggregate a worker's performance history across all platforms, creating a far more holistic and trustworthy measure of skill and reliability than any single platform's rating system.
    - **Settlement Protocol**: A standardised process for triggering payments upon the completion of work or milestones. This layer could integrate directly with financial protocols like UPI, enabling smart contracts where, for instance, a verified "work complete" signal on UWI automatically initiates a payment.

- **A Neutral Governance Body:** Just as the National Payments Corporation of India (NPCI) oversees UPI, a neutral, non-profit, or quasi-governmental body would be essential. This entity would be responsible for maintaining the protocol's integrity, ensuring data security, resolving disputes, and guiding its evolution.

**Standardising the Language of Work**

For the protocol to function, it must standardise the information that defines both workers and the work itself.

- **The Worker Profile Schema:** This would be a comprehensive digital representation of a worker's professional life, containing structured fields for verified identity, skills (using a standardised taxonomy), credentials (from verified issuers like universities and certification bodies, or eGov Foundation's DIVOC credentialing module (eGov Foundation, 2022), a complete work history, and the aggregated reputation data.
- **The Work Packet Schema:** This would be the universal job description. It would include a unique ID for the task, verified employer information, a detailed scope of work, the specific skills required (using the same taxonomy), and clear terms for logistics (remote/on-site, duration) and compensation (type, amount, payment terms).

**An Ecosystem of Innovation: Applications Built on UWI**

The true power of UWI would be realised in the explosion of applications built on top of it. By providing the underlying infrastructure, UWI would dramatically lower the barrier to entry for innovators. Potential applications include:

- **Creating a Safety Net:** A primary obstacle for governments aiming to provide a safety net for gig workers is the lack of a unified system to identify this transient workforce and track their work across disparate platforms. UWI would solve this challenge by acting as the digital public infrastructure, or "rails," upon which social security schemes can run. The UWI-ID would provide a single, verifiable identity for every worker. Also, because every "Work Packet" and payment is recorded on this protocol, UWI creates a real-time, trustworthy ledger of a worker's activity and income. This verified ledger could then directly integrate with parallel digital public infrastructures like the Unified Benefits Interface (UBI), creating an end-to-end system for contribution and benefit delivery (Piramal Foundation, 2025).

- This would allow a government to seamlessly implement laws like the Code on Social Security, 2020, by enabling automated, transaction-based deductions for welfare funds (e.g., health insurance, pension contributions) directly from the platform at the point of payment. This makes the collection of contributions and the verification of eligibility for benefits transparent, efficient, and directly linked to a worker's actual economic activity, turning the aspiration of a social safety net into an operational reality.
- **Hyper-Specialised Talent Platforms**: Imagine a platform designed exclusively for hiring certified plumbers or electricians for household repairs. Another could focus on connecting licensed commercial drivers with logistics companies for short-haul routes. These platforms wouldn't need to build a user base from scratch; they would simply plug into the UWI network and focus on creating the best possible user experience for their niche.
- **Financial Services**: A delivery driver with a strong, positive work history on their UWI-ID could get a competitive loan for a new vehicle. A freelance construction worker could access insurance products that cover them on a per-project basis. With access to verified income streams via the UWI protocol, fintech companies could offer a range of tailored financial products to this underserved segment.
- **Intelligent Workforce Management for Enterprises**: A large construction firm could use an internal tool that plugs into UWI to find certified welders or forklift operators for a specific phase of a project. A major event management company could seamlessly hire and manage temporary crews for setup and teardown, all verified through the UWI protocol.
- **Dynamic Government Skilling Initiatives**: Governments could analyse real-time UWI data to identify emerging skill gaps in the economy. They could then target specific training programs to citizens who need them most and instantly issue a verified digital credential to their UWI-ID upon completion.

The Unified Workers Interface (UWI) represents a shift from the fragmented, siloed labour markets of today to a connected and dynamic future. By creating a digital public infrastructure for work, UWI offers a blueprint for a system that empowers gig workers & individuals with ownership of their professional identity, provides employers with frictionless access to talent, and enables a new wave of innovation in the technology of work.

## 5.6. Design/Dark Patterns for Gig Work

Design Patterns relate to the systemic design of the platform's operations and business model. Addressing them requires changes to platform policies, algorithmic logic, and potentially regulatory intervention. Dark Patterns shall mean "any practices or deceptive design pattern using user interface or user experience interactions on any platform that is designed to mislead or trick users to do something they originally did not intend or want to do, by subverting or impairing the consumer autonomy, decision making or choice, amounting to misleading advertisement or unfair trade practice or violation of consumer rights" (Central Consumer Protection Authority, 2023).

| Type | Description |
|---|---|
| **Design Pattern** | **Algorithmic Opacity**<br>Gig workers don't know how decisions are made by the platform's algorithm, creating uncertainty (Chen et al., 2015; Rosenblat & Stark, 2016). |
| **Design Pattern** | **Hidden Costs / Offloading of Risk**<br>Platforms are not upfront about operational costs that workers must bear (e.g., vehicle maintenance, fuel) (Ravenelle, 2019). |
| **Design Pattern** | **Moving Targets**<br>Platforms are known to alter performance targets or incentive thresholds unpredictably and without transparency, frequently shifting the requirements to achieve promised rewards (Vasudevan and Chan, 2022). |
| **Design Pattern** | **Biased Rating Systems**<br>Customer ratings are used as a primary disciplinary tool, often penalising workers for factors beyond their control (e.g., restaurant delays, customer mood (Lee et al., 2015). |

| Type | Description |
| --- | --- |
| **Design + Dark Pattern** | **Unpredictable Scheduling**<br>Despite the promise of flexibility, work availability can be highly volatile and algorithmically controlled, making it difficult for workers to plan or secure stable income (Reynolds et al., 2024). |
| **Design + Dark Pattern** | **Excessive Monitoring**<br>Intrusive, constant surveillance of workers through their devices, tracking location, activity, speed, and communication for performance management (Rosenblat & Stark, 2016). |
| **Design + Dark Pattern** | **Unfair Deactivation**<br>Workers are permanently or temporarily removed from the platform without a clear reason, evidence, or a meaningful process for appeal (Heeks et al., 2021). |
| **Dark Pattern** | **False Urgency**<br>A sense of urgency is created (e.g., countdown timers) to compel workers to accept a gig immediately without full consideration (Lee et al., 2015). |
| **Dark Pattern** | **Wage Concealment**<br>The full breakdown of how pay is calculated is obscured from the worker, often until after a job is completed (Rosenblat & Stark, 2016). |

### Principles to Address these Patterns

We believe that clear UI/UX, product, and governance standards should be established to reduce harms from design and dark patterns in gig economy apps. The following principles may form the ethical foundation for a platform that treats its workers as valued partners.

1. **Transparency in Practice**: Workers have the right to understand how they earn and how work is assigned. All inputs, formulas, and policy changes must be visible, versioned, and explainable in plain language. This includes providing upfront information on every gig and clear, itemized receipts for every completed task.
2. **Worker Agency & Control**: Workers have the right to make informed decisions without psychological manipulation or surrendering excessive data. Consent for data collection must be granular and progressive, with clear scopes, purposes, and opt-outs. The user interface must be neutral, avoiding coercive elements like aggressive countdowns or shaming language.
3. **Fairness in Feedback**: Workers have the right to be evaluated on factors within their control. Rating systems must be intelligently designed to automatically adjust to external conditions like traffic, weather, or restaurant delays that are beyond the worker's control.
4. **Predictability & Stability**: Workers have the right to a stable and predictable work environment that allows for financial planning. This is achieved through tools like demand forecasting, the ability to book work slots, and establishing minimum earnings floors where feasible.
5. **Due Process & Respect**: Workers have the right to a clear explanation and a fair hearing before losing their livelihood. All enforcement actions must be backed by accessible evidence, with a clear, guided appeals process that has a time-bound resolution.

## Design/ Dark Pattern Remediation

| Pattern | Remediation |
|---|---|
| **Algorithmic Opacity** | • **Explainable AI:** Provide "explain-as-you-go" justifications for offers directly in the app (e.g., "This offer was based on proximity and your 7-day ETA reliability").<br>• **Pay Transparency:** Include a "Why this payout?" explainer for every gig with a full formula breakdown.<br>• **Transparent Changes:** Maintain an in-app, versioned change log for all policies, highlighting what changed and who is affected.<br>• **Audits**: Conduct independent algorithmic audits, in collaboration with research institutions, to check for fairness and disparate impact. |
| **Hidden Costs / Offloading of Risk** | • **Net Earnings Calculators:** Display net-earnings calculators with editable, localised cost assumptions (fuel, maintenance, taxes) to show a realistic take-home pay estimate.<br>• **Risk Mitigation:** Shift risk where possible by offering a minimum per-KM floor indexed to local fuel prices. |
| **Moving Targets (Incentives)** | • **Fixed Terms:** Treat incentives like financial products by providing "term sheets" with fixed time windows and clear eligibility rules.<br>• **Advance Notice:** Lock all changes with a 7-14 day advance notice period (unless for fraud/safety), communicated via an in-app countdown.<br>• **Progress Tracking:** Provide personalised progress trackers with "what-if" calculators to help workers strategise. |

```
WEEKLY BOOST v1.8 (Effective: 02-08 Sep)  [See v1.7]
• 40 trips → +₹600  • 60 trips → +₹1,400  (Cap ₹1,400)
Notice: Next change in 6 days.                  [Changelog]
Your progress: 18/40  [What if I do +5 today? → ETA + payout]
```

| Pattern | Remediation |
|---|---|
| **Biased Rating Systems** | • **Contextual Normalisation**: Use multi-source signals (traffic feeds, weather, restaurant delay flags) to automatically normalise ratings and remove penalties for external factors.<br>• **Coaching over Punishment**: Replace binary demerits with constructive coaching nudges and forgiveness windows (e.g., first 2 flagged events per month are not penalized).<br>• **Context Tags**: Allow workers to add context tags (e.g., "Customer unresponsive") to a rating, which can trigger a review or remove a penalty. |

```
Today's impact: ★★★★☆ (4.4)
Context tags auto-applied: [Restaurant delay 18m] [Heavy rain]
Penalty removed: Late arrival adjusted by +0.2 due to traffic event.
Next step: Keep customer informed → Template: "Delay at kitchen…".
```

| | |
|---|---|
| **Unpredictable Scheduling** | • **Slot Booking**: Offer the ability to book work slots in advance, offer soft holds for peak times.<br>• **Demand Forecasting**: Publish supply-demand heatmaps and forecast windows for the next 24-72 hours to help workers plan their time.<br>• **Minimum Earnings Floor**: Show-up fee, honour holds or compensation when cancelled. |

```
Book Slots (Thu)
08-10  Medium (₹120/hr floor)   10-12  High (₹160/hr floor)
Heatmap (Next 3h):  ███ High  ██ Med  █ Low   [View forecast]
If floor not met → Top-up paid within 24h.  [Policy]
```

**Design/ Dark Pattern Remediation**

| Pattern | Typical UI Patterns | Remediation |
|---|---|---|
| **False Urgency** | • Gig cards with countdown timers (e.g., 5–10 seconds), pulsing colours, or loss-framing copy ("Hurry! Surge ends!"). | • Replace countdowns with a "soft hold" of at least 30-60 seconds, with an option to extend once.<br>• Use neutral colours and copy; remove alarming animations. |

```
┌──────────────────────────────────────────────────────────────┐
| Pickup → Drop | 5.2 km | Est. 22-28 min                       |
| PAY: ₹168-₹192  [View Breakdown]  [Why this offer?]          |
|                                                              |
| [ACCEPT]   [PASS]    Hold: 00:40  ( Need more time? [+30s] ) |
└──────────────────────────────────────────────────────────────┘
Breakdown (expanded): Base ₹90 + Distance ₹48 + Surge x1.2 +
Cancellation ₹0 – Fees ₹6. Tip: post-delivery, visible if added.
Why this offer? Proximity, ETA reliability (7-day), past zone prefs.
```

| | | |
|---|---|---|
| **Wage Concealment** | • "Accept to see final pay" screens.<br>• Pay shown only post-completion or after timers expire.<br>• Tip policy unclear. | • The offer card must show the full pay formula, distance, and estimated time upfront.<br>• A "View Breakdown" button must be present and viewable before acceptance.<br>• Receipts UI post-trip that reconcile estimate vs realised pay, with dispute button next to each component. |

```
Costs Preset: Fuel ₹110/L • Mileage 35 km/L • Maint. ₹2.5/km
Trip est. net: ₹168 – ₹(fuel 5.2/35*110=₹16) – Maint ₹13 = **₹139**
[Edit assumptions]  [Save as default]
After trip: Est ₹139 → Realised ₹142  [Dispute]
```

| | | |
|---|---|---|
| **Excessive Monitoring** | • Hidden toggles for background location.<br>• Default-on call recording/always-on mic permissions bundled with essential permissions.<br>• “Functionally impossible to work” if strict tracking is not accepted, without clear scope/purpose. | • Use separate toggles for each permission (e.g., "Location while using" vs. "Location always"), each with an explicit scope and rationale.<br>• Provide a screen showing what data was collected in the last 24 hours (GPS pings, selfie checks), for what purpose, and for how long it's retained. Include buttons to delete eligible data or request a review.<br>• Offer a visible "Privacy Break" mode that clearly suspends non-essential tracking.<br>• No critical app function is blocked by a single bundled permission; every permission has an independent toggle and explanation. |

```
Permissions
[ ] Location (while using)   [ ] Location (always) (for SOS only)
[ ] Selfie checks (biometric) (fraud prevention; stored 90 days)
[ ] Call metadata (safety) (no audio; stored 30 days)

Data Ledger (last 24h)
• 312 GPS pings (routing)   • 1 selfie (shift start)   • 2 call logs
[Delete eligible]  [Request review]  [Privacy break: 15:00–15:20]
```

| | | |
|---|---|---|
| **Unfair Deactivation** | • One-line "policy violation" banners with no evidence.<br>• Appeals hidden under multiple taps; deadlines unclear; status opaque. | • A deactivation notice must be presented as a "Deactivation Card" that includes the specific policy clause violated, timestamped evidence (with sensitive data blurred), and the impact (e.g., temporary suspension).<br>• The notice must feature a prominent "Start Appeal" button that initiates a guided flow. This flow should include a checklist of required proofs, an upload tool, and a real-time status tracker (e.g., "Submitted," "Under Review," "Decision by [Date]").<br>• Implement a UI that shows a progression from warning to temporary timeout to deactivation, with each step providing advice on how to avoid escalation.<br>• No deactivation notice is served without viewable evidence and a prominent appeal button on the same screen. The appeals process must have a stated, time-bound resolution window. |

```
⚠ Account Temporarily Suspended (Policy 4.2: Fraudulent activity)
Evidence: Order #A7821 – mismatch on selfie @ 14:22 (image shown)
Impact: Temp suspension (72h) – earnings unaffected.
[Start Appeal]  Deadline: 48h   ETA resolution: 24–72h
Appeal steps: (1) Upload selfie now  (2) Govt ID  (3) Explain context
Status: Submitted → Under review (agent: Priya) → Decision by 04 Sep
```

# Appendix: List of Stakeholder Interviewees

- Aditi Surie, Assistant Professor, School of Human Development & Lead – Academics & Research, Indian Institute for Human Settlements
- Amal Sivaji, Co-Founder, Draavi
- Ambika Tandon, PhD student, University of Cambridge
- Amit Jain, Chief Operating Officer, FluxGen
- Balaji Parthasarathy, Center for Information Technology and Public Policy, International Institute of Information Technology, Bangalore, and Principal Investigator, Fairwork India
- Dharmendra Kumar, Founding Secretary, Janpahal
- Gayatri Nair, Assistant Professor of Sociology, Indraprastha Institute of Information Technology, Delhi
- Hari Menon, Co-Founder and CEO, BigBasket
- Lavanya Singh, Founder & CEO, Kaio
- Priyam Vadaliya, Manager, Aapti Institute
- Rakshita Swamy, Visiting Faculty, National Law School of India University, Bengaluru & Head, Social Accountability Forum for Action and Research (SAFAR)
- Rameesh Kailasam, CEO and President, IndiaTech.org
- Ravi Annadurai, CEO, Workaroo
- Rikta Krishnaswamy, All India Gig Workers Union
- Ritvik Gupta, Associate, Aapti Institute
- Senior Executive, Platform Company 1
- Senior Executive, Platform Company 2
- Shobhit S. Lead - Research & Policy Engagement, IT for Change
- Shourya Roy, Senior Vice President and Head of Data Science, Zepto
- Vaibhav Raaj, National Project Officer, India, International Labour Organization (ILO)
- Vedant Sharma, All India Gig Workers Union

# References

- Acemoglu, D., & Restrepo, P. (2019). Automation and new tasks: How technology displaces and reinstates labor. Journal of Economic Perspectives, 33(2), 3-30. https://www.aeaweb.org/articles?id=10.1257/jep.33.2.3
- Adams-Prassl, Jeremias (2019). What if Your Boss Was an Algorithm? The Rise of Artificial Intelligence at Work. [2019] 41(1) Comparative Labor Law & Policy Journal 123, Available at SSRN: https://ssrn.com/abstract=3661151
- All India Gig Workers' Union (AIGWU). (2023). Your economy, our livelihoods: All India Gig Workers' Union critique on NITI Aayog's report on the platform economy. Centre for Internet and Society. https://cis-india.org/raw/files/your-economy-our-livelihoods.pdf.
- Assocham - Primus Partners, 2021. The Rising Gig Economy of India. https://www.primuspartners.in/docs/documents/The%20Rising%20GIG%20Economy%20Of%20India-09-21-06.pdf.
- Autor, D. H. (2015). Why are there still so many jobs? The history and future of workplace automation. Journal of Economic Perspectives, 29(3), 3-30. https://www.aeaweb.org/articles?id=10.1257/jep.29.3.3
- Autor, D. H., Levy, F., & Murnane, R. J. (2003). The skill content of recent technological change: An empirical exploration. Quarterly Journal of Economics, 118(4), 1279-1333. https://academic.oup.com/qje/article-abstract/118/4/1279/1925105
- AZB & Partners. (2023, May 9). Social Security Code provisions pertaining to pension on higher wages notified. AZB & Partners. https://www.azbpartners.com/bank/social-security-code-provisions-pertaining-to-pension-on-higher-wages-notified/
- Barocas, Solon and Selbst, Andrew D., Big Data's Disparate Impact (2016). 104 California Law Review 671 (2016), Available at SSRN: https://ssrn.com/abstract=2477899 or http://dx.doi.org/10.2139/ssrn.2477899
- Berg, J. (2024, October 9). Minimizing the negative effects of AI-induced technological unemployment. International Labour Organization. https://www.ilo.org/resource/article/minimizing-negative-effects-ai-induced-technological-unemployment
- Boston Consulting Group & Michael & Susan Dell Foundation. (2021). Unlocking the Potential of the Gig Economy in India. Boston Consulting Group. https://media-publications.bcg.com/India-Gig-Economy-Report.pdf
- Bresnahan, T. F., Brynjolfsson, E., & Hitt, L. M. (2002). Information technology, workplace organization, and the demand for skilled labor: Firm-level evidence. Quarterly Journal of Economics, 117(1), 339-376. https://academic.oup.com/qje/article-abstract/117/1/339/1851770

- Brynjolfsson, E., Rock, D., & Syverson, C. (2017). Artificial intelligence and the modern productivity paradox: A clash of expectations and statistics (NBER Working Paper No. 24001). National Bureau of Economic Research. https://www.nber.org/system/files/working_papers/w24001/w24001.pdf
- Central Consumer Protection Authority, 2023. Guidelines for prevention and regulation of dark patterns, 2023. Department of Consumer Affairs, Ministry of Consumer Affairs, Food & Public Distribution, Government of India.
- Chen, L., Mislove, A., & Wilson, C. (2015). Peeking Beneath the Hood of Uber. Proceedings of the 2015 ACM Conference on Internet Measurement Conference - IMC '15. https://dl.acm.org/doi/10.1145/2815675.2815681
- Deepak, M., Mansi, K., Aarti, R., Krithika, R., & Mayank, M. (2024). State of India's Digital Economy (SIDE) Report, 2024. Indian Council for Research on International Economic Relations (ICRIER). https://icrier.org/pdf/State_of_India_Digital_Economy_Report_2024.pdf
- eGov Foundation, 2022. DIVOC Modules. https://divoc.egov.org.in/divoc-modules/.
- Fairwork India. (2023). Fairwork India Ratings 2023: Labour Standards in the Platform Economy. Fairwork Project. https://fair.work/en/fw/publications/fairwork-india-ratings-2023-labour-standards-in-the-platform-economy/
- Galtung, J. (1969). Violence, Peace, and Peace Research. Journal of Peace Research, 6(3), 167–191. https://doi.org/10.1177/002234336900600301
- Heeks, R., Graham, M., Mungai, P., Van Belle, J. P., & Woodcock, J. (2021). Systematic evaluation of gig work against decent work standards: The development and application of the Fairwork framework. The Information Society, 37(5), 267–286. https://doi.org/10.1080/01972243.2021.1942356.
- IDinsight, (2025, May 6), Economic lives of digital platform gig workers: Case of delivery drivers in India. IDinsight. https://www.idinsight.org/publication/economic-lives-of-digital-platform-gig-workers-india
- International Labour Conference (2025), Resolution to place on the agenda of the next ordinary session of the Conference an item entitled "Decent work in the platform economy". https://www.ilo.org/sites/default/files/2025-07/ILC113-Resolution%20V-%5BRELMEETINGS-250612-016%5D-EN.pdf.
- International Labour Organization, 2021. World Employment and Social Outlook 2021: The role of digital labour platforms in transforming the world of work. Geneva: ILO. https://www.ilo.org/publications/flagship-reports/role-digital-labour-platforms-transforming-world-work
- International Labour Organization, 2023. Generative AI and jobs: A global analysis of potential effects on job quantity and quality (Working Paper No. 96). Geneva: ILO. https://www.ilo.org/publications/generative-ai-and-jobs-global-analysis-potential-effects-job-quantity-and

- International Labour Organization. (2024, January 31). Realizing decent work in the platform economy (Report V(1), International Labour Conference, 113th Sess.). International Labour Office. https://www.ilo.org/resource/conference-paper/ilc/113/realizing-decent-work-platform-economy
- International Labour Organization. (2025, August 15). Decent work in the platform economy (Draft Convention and Recommendation, International Labour Conference, 114th Sess.). International Labour Office. https://www.ilo.org/resource/conference-paper/ilc/ilc114/decent-work-platform-economy
- Jain, A. (2023, October 30). Minimum wage policies absent for most of India's platform gig workers: Report. Deccan Herald. https://www.deccanherald.com/business/minimum-wage-policies-absent-for-most-of-indias-platform-gig-workers-report-2748314.
- Janpahal (2024). The RIGHTS Survey - Report on a nationwide survey of platform workers in India. https://www.janpahal.com/_files/ugd/1cfdd7_f9ce99577d974faa9ef28c9fceb23761.pdf
- Kellogg, K. C., Valentine, M. A., & Christin, A. (2020). Algorithms at work: The new contested terrain of control. Academy of Management Annals, 14(1), 366–410. https://journals.aom.org/doi/10.5465/annals.2018.0174
- Lee, M. K., Kusbit, D., Metsky, E., & Dabbish, L. (2015). Working with Machines. Proceedings of the 33rd Annual ACM Conference on Human Factors in Computing Systems - CHI '15, 1603–1612. https://doi.org/10.1145/2702123.2702548.
- Madhulika, T., Sanjay, S., & Vaidya, E. (2025, March). Digital bosses: The role of algorithms in the modern workplace. National Law School of India University & IT for Change. https://itforchange.net/sites/default/files/add/Digital%20Bosses%20The%20Role%20of%20Algorithms%20in%20the%20Modern%20Workplace.pdf
- Ministry of Information & Broadcasting, 2024. India's Digital Revolution: Transforming Infrastructure, Governance, and Public Services. (2024). https://www.pib.gov.in/PressReleaseIframePage.aspx?PRID=2082144.
- Ministry of Labour & Employment. (2025a, March 8). Ministry of Labour and Employment Urges Platform Workers to Register on e-Shram Portal for Formal Recognition and Access to AB-PMJAY Benefits. Press Information Bureau. https://www.pib.gov.in/PressReleasePage.aspx?PRID=2109421.
- Ministry of Labour & Employment. (2025b, August 11). e-SHRAM platform enhancements. Press Information Bureau. https://www.pib.gov.in/PressReleasePage.aspx?PRID=2155172.
- Mohlmann, M., & Zalmanson, L. (2017). Hands on the Wheel: Navigating Algorithmic Management and Uber Drivers' Autonomy. ICIS 2017 Proceedings. https://aisel.aisnet.org/icis2017/DigitalPlatforms/Presentations/3.

- Moneycontrol, 2025. Uber pilots in-app AI task program for drivers in India. https://www.moneycontrol.com/artificial-intelligence/uber-pilots-in-app-ai-task-program-for-drivers-in-india-article-13522812.html
- National Council of Applied Economic Research, 2023. Socio-economic Impact Assessment of Food Delivery Platform Workers, https://ncaer.org/wp-content/uploads/2023/08/NCAER_Report_Platform_Workers_August_28_2023.pdf
- NITI Aayog. (2022). India's booming gig and platform economy: Perspectives and recommendations on the future of work. Government of India. https://www.niti.gov.in/sites/default/files/2022-06/25th_June_Final_Report_27062022.pdf
- PAIGAM, University of Pennsylvania, 2024. PRISONERS ON WHEELS? Report on Working and Living Conditions of App-based workers in India. https://tgpwu.org/wp-content/uploads/2024/03/Report-Final-Print-1.pdf.
- Pan, J. C.-H., & Wu, M.-H. (2012). Throughput analysis for order picking system with multiple pickers and aisle congestion considerations. Computers & Operations Research, 39(7), 1661–1672. https://doi.org/10.1016/j.cor.2011.09.022.
- Piramal Foundation, 2025. UBI is launched: An open network for inclusive welfare. https://piramalfoundation.org/ubi-is-launched-an-open-network-for-inclusive-welfare.
- Ramaswamy, Venkat and Narayanan, Krishnan (2025). The Co-Intelligence Revolution: How Humans and AI Co-Create New Value. Penguin Business.
- Ravenelle, A. J. (2019). Hustle and gig : struggling and surviving in the sharing economy. University Of California Press.
- Rawls, J. (1971). A Theory of Justice: Original Edition. Harvard University Press. https://doi.org/10.2307/j.ctvjf9z6v
- Reynolds, J., Quintero, D. F. P., Aguilar, J., & Kincaid, R. (2024). Work schedules, finances, and freedom: Work schedule fit and platform dependence among gig workers on Amazon's Mechanical Turk platform. Socius, 10. https://doi.org/10.1177/23780231241286933
- Rosenblat, A., & Stark, L. (2016). Algorithmic labor and information asymmetries: A case study of Uber's drivers. International Journal of Communication, 10, 3758–3784. https://papers.ssrn.com/sol3/papers.cfm?abstract_id=2686227
- Sakshi Sadashiv K. (2025, January 2). Why Gig Workers Are Calling for Algorithmic Data Transparency. MEDIANAMA. https://www.medianama.com/2025/01/223-why-the-gig-workers-union-is-calling-for-algorithmic-data-transparency/.
- Sen, A. (1999). Development as freedom. Oxford University Press.
- Sen, S., Chatterjee, S., Agarwal, S., & Sharma, A. (2023). Centring rights in the platform workplace: Understanding legal frameworks around algorithmic management and upholding worker rights (IT for Change; Centre for Labour Studies, NLSIU). https://itforchange.net/sites/default/files/2491/Centring%20Rights%20in%20the%20Platform%20Workplace_ITfC_0.pdf.

- The Bihar Platform Based Gig Workers (Registration, Safety and Welfare), Act, 2025, PRS Legislative Research. https://prsindia.org/files/bills_acts/acts_states/bihar/2025/Act8of2025BR.pdf
- The Code on Social Security (2020). Ministry of Law & Justice. https://labour.gov.in/sites/default/files/ss_code_gazette.pdf
- The Draft Telangana Gig and Platform Workers (Registration, Social Security, and Welfare) Bill, 2025. https://prsindia.org/bills/states/the-draft-telangana-gig-and-platform-workers-registration-social-security-and-welfare-bill-2025
- The Hindu. (2021, December). An estimated 12.2 crore Indians lost their jobs during the coronavirus lockdown in April: CMIE. https://www.thehindu.com/data/data-over-12-crore-indians-lost-their-jobs-during-the-coronavirus-lockdown-in-april/article61660110.ece
- The Draft Jharkhand Platform Based Gig Workers (Registration and Welfare) Bill, 2024. https://shramadhan.jharkhand.gov.in/ftp/WebAdmin/documents/Gig-worker-social-security.pdf
- The Karnataka Platform Based Gig Workers (Social Security and Welfare) Bill, 2025. https://kla.kar.nic.in/council/house/bills/156/31E.pdf
- The Rajasthan Platform Based Gig Workers (Registration and Welfare) Act, 2023, Act No. 29 of 2023. https://shramadhan.jharkhand.gov.in/ftp/WebAdmin/documents/Gig-worker-social-security.pdf
- Vasudevan, K., & Chan, N. K. (2022). Gamification and work games: Examining consent and resistance among Uber drivers. New Media & Society, 24(4), 866-886. https://doi.org/10.1177/14614448221079028
- Vasudevan, V. (2025, July). OTP Please: Online buyers, sellers and gig workers in South Asia. India Penguin.
- V.V. Giri National Labour Institute, 2025. Gig and Platform Workers: Vision 2027, Dhanya M.B., NLI Research Studies Series No. 173/2025. https://www.vvgnli.gov.in/sites/default/files/173-2025%20Dhanya%20MB.pdf
- Woodcock, J., & Graham, M. (2020). The Gig Economy: A Critical Introduction. http://acdc2007.free.fr/woodcock2020.pdf

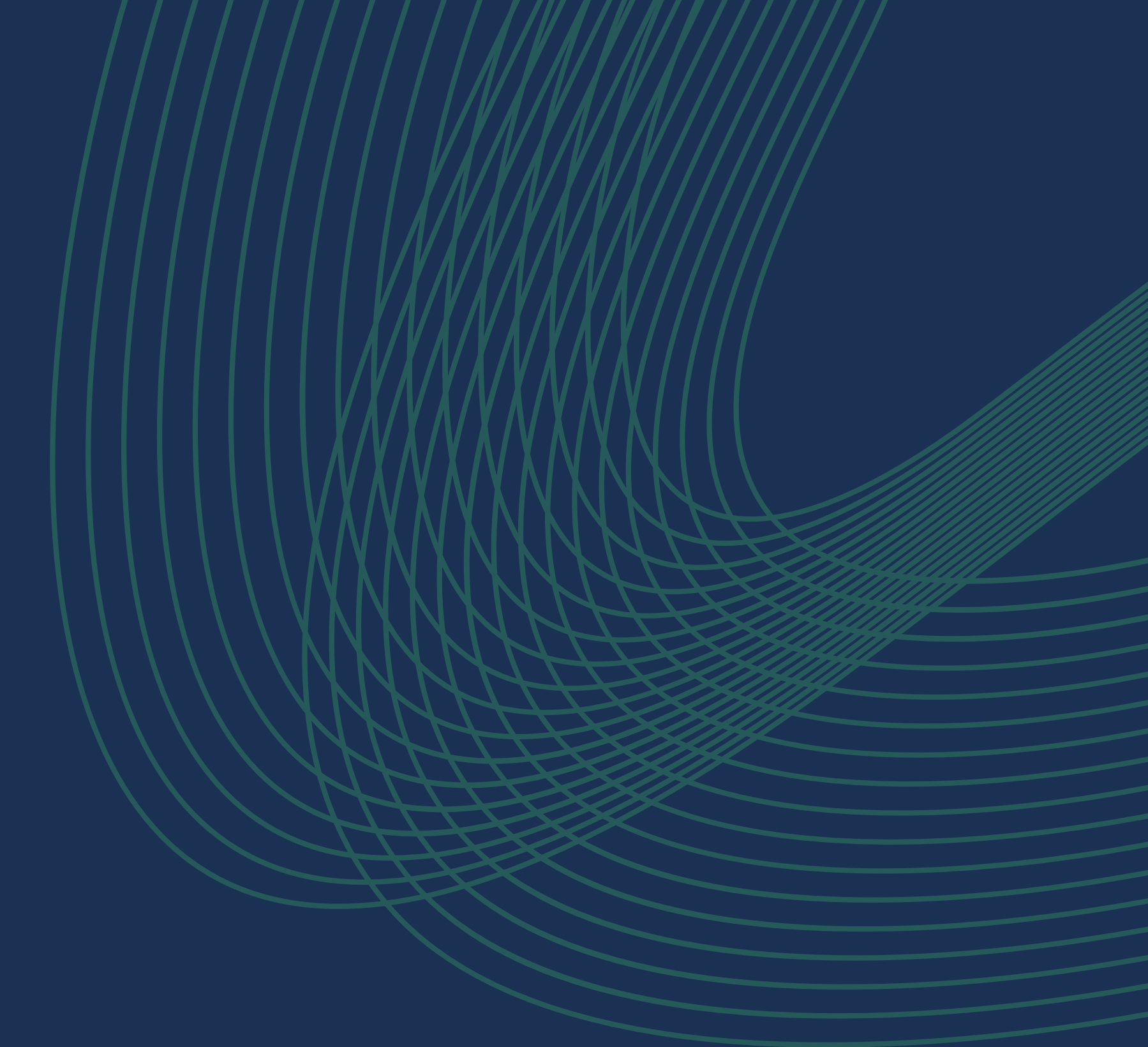

# The Algorithmic-Human Manager: AI, Apps, and Workers in the Indian Gig Economy

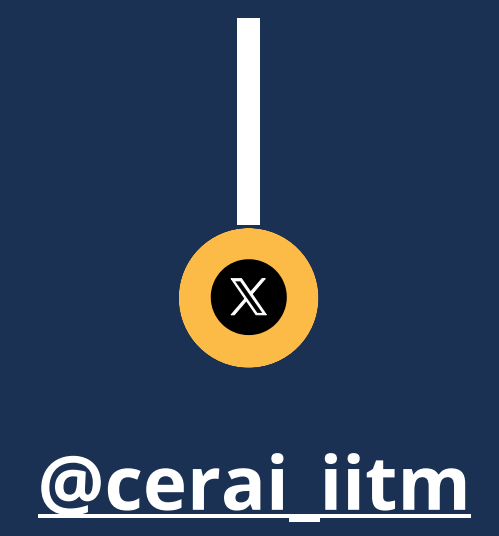
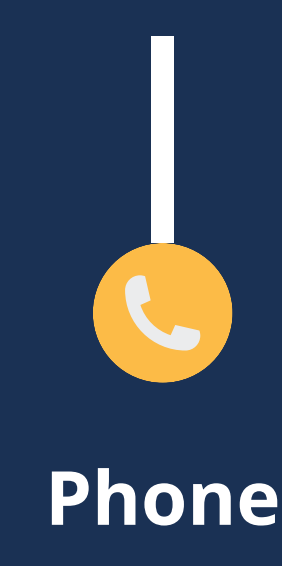
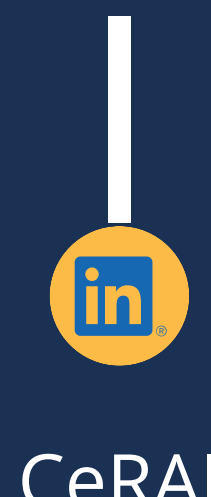

@cerai_iitm

**Phone**

CeRAI

+91-44-22578980

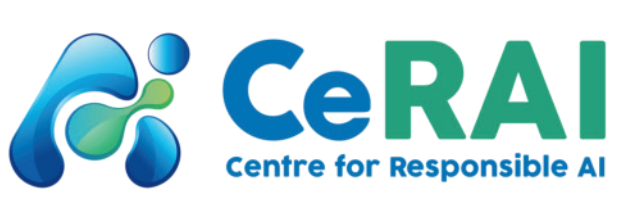

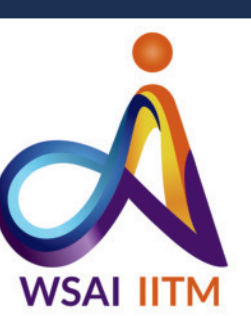

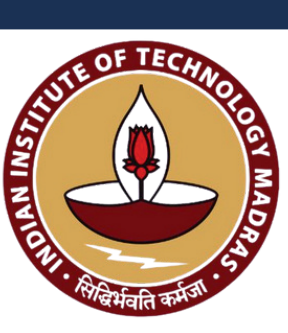